\documentclass[twocolumn]{aastex701}

\usepackage{amsmath}
\usepackage{physics}
\usepackage{bbold}
\usepackage{soul}

\graphicspath{{./}{figures/}}

\begin{document}

\title{The Impact of Cosmic Ray Transport on the $\gamma$-Ray Luminosity of Diffuse Gas}

\author[0000-0003-4776-940X]{Roark Habegger}
\affiliation{University of Wisconsin-Madison Astronomy Department }
\email[show]{rhabegger@wisc.edu}

\author[0009-0002-2669-9908]{Mateusz Ruszkowski}
\affiliation{University of Michigan Department of Astronomy}
\email{rhabegger@wisc.edu}

\author[0000-0003-4821-713X]{Ellen G. Zweibel}
\affiliation{University of Wisconsin-Madison Astronomy Department }
\affiliation{University of Wisconsin-Madison Physics Department }
\email{rhabegger@wisc.edu}

\begin{abstract}

Observations of $\gamma$-rays from diffuse gas provide the opportunity to study the distribution of high energy particles in different astrophysical environments. In the circumgalactic medium (CGM) and the intracluster medium (ICM), it is expected that relativistic cosmic rays collide with thermal particles and produce $\gamma$-rays through pion decay. The $\gamma$-ray luminosity of a plasma depends on where cosmic rays are: if they are in denser gas, they produce more $\gamma$-rays. In this work, we study how different cosmic-ray transport mechanisms impact the $\gamma$-ray luminosity of a turbulent, multiphase medium formed from an initially diffuse medium. Two quantities set the luminosity: the average cosmic-ray energy density and the correlation of cosmic-ray energy and gas density. Overall, cosmic rays must escape cold dense regions in order to produce less $\gamma$-ray emission and be consistent with observations. Our simulations with fast transport mechanisms (either diffusion or streaming) are degenerate: they each produce a lower $\gamma$-ray luminosity than slow transport simulations by two orders of magnitude. This result means that fast transport (particularly in dense clumps) is necessary for simulations to agree with the dearth of observations of $\gamma$-ray emission from diffuse gas like the CGM and ICM. We also show the significant difference in luminosity is the result of cosmic-ray reacceleration. This reacceleration is different from the turbulent reacceleration described by Ptuskin (1988). Instead, condensing, cold clouds drive a significant increase in the average cosmic-ray energy and, as a result, the $\gamma$-ray luminosity.

\end{abstract}


\section{Introduction} \label{sec:intro}

The most common cosmic rays are low-energy protons ($\sim1-10\,\mathrm{GeV}$). They make up a majority of the total cosmic-ray energy density in our local interstellar medium (ISM) \citep{2016ApJ...829....8C,2019PhRvD..99j3023E}. While detectors like Voyager and AMS directly detect them in the solar neighborhood and provide this estimate, it is challenging to analyze low-energy cosmic rays outside the nearby ISM. While ultra-high energy cosmic rays can ballistically reach us from other galaxies \citep{2023Sci...382..903T}, they provide no information about the low-energy cosmic rays in that galaxy. 

Instead, studies of cosmic rays in the diffuse circumgalactic medium (CGM) and intracluster medium (ICM) rely on observing the radiation emitted as a result of cosmic rays interacting with thermal plasma (see recent reviews by \citep{2023Galax..11...86O} and \citep{2023Ruszkowski}). As cosmic ray protons pass through thermal plasma, they can hit other nuclei and trigger a hadronic interaction which produces pions \citep{2007Enslin,2008MNRAS.384..251G}. Some of these pions decay into $\gamma$-rays, which can propagate through the gas and eventually reach us. Only cosmic rays with an energy above a threshold energy $E_\mathrm{th} = 1.22\, \mathrm{GeV}$ in the rest frame of the observer can trigger this interaction, but the common GeV protons meet this criterion. Since they are the most common, they dominate this $\gamma$-ray emission. As a result, observations of $\gamma$-ray emission can be used to constrain the total cosmic-ray energy density in our own galaxy, as well as in galaxies and galaxy clusters far away from us.

However, there is a problem with directly calculating the cosmic-ray energy density from the $\gamma$-ray emission. The $\gamma$-ray luminosity $L_\gamma$ is determined by the product of the gas density and the cosmic-ray energy density. The average of a product $(\langle f g\rangle)$can deviate significantly from the product of averages $(\langle f \rangle \langle g\rangle)$ for a multiphase gas \citep{2013ApJ...779...12B}. When averaging over a volume of multiphase gas, we separate the total luminosity into three quantities: the average cosmic-ray energy density, the average gas density, and the correlation of cosmic-ray energy density and gas density. The correlation of the cosmic-ray energy density and gas density (quantified in Section \ref{sec:phys:corr}) depends on the thermodynamic, magnetic, and turbulent properties of the gas. But primarily, the correlation depends on the cosmic-ray transport in the volume. 

The interaction of cosmic rays with a turbulent medium is a timely problem, as Ptuskin's acceleration of cosmic rays by long-wavelength turbulence \citep{1988SvAL...14..255P} is being reexamined with modern computational techniques \citep{2022ApJ...941...65B,2023ApJ...955...64B,2024ApJ...974...17H,2025arXiv250603768S}. This acceleration could increase the cosmic-ray energy, and therefore the $\gamma$-ray luminosity. We will review the expected growth rates, but the important takeaway from previous works is the mechanism should be most optimal and have its largest impact in the CGM \citep{2023ApJ...955...64B}.

In this work, we examine how cosmic-ray transport influences $\gamma$-ray luminosity of diffuse, magnetized, and turbulent plasma. We show that transport influences the $\gamma$-ray luminosity through two mechanisms: (1) increasing the average cosmic-ray energy density through turbulent reacceleration, and (2) changing the correlation of cosmic-ray energy density and gas density. If the cosmic rays do not escape dense gas, then the $\gamma$-ray luminosity rapidly increases as a result of both cosmic-ray acceleration and correlation with dense gas. In that case, our simulations disagree with null detections of diffuse $\gamma$-ray emission from diffuse gas (e.g., in galaxy clusters). Therefore, we can place limits on the cosmic-ray transport in diffuse media such as the CGM and ICM. We find that the diffusion coefficient must be $ \gtrsim  10^{30}\,\mathrm{cm}^2 \mathrm{s}^{-1}$ or that streaming (self-confinement model of cosmic-ray transport) must dominate transport in diffuse gas. We also find reacceleration by condensing, cold clouds is more effective at reacceleration of cosmic rays than turbulent reacceleration in a single-phase medium. 

This paper proceeds first with background on the calculation of the $\gamma$-ray luminosity, turbulent acceleration, and correlation in Section \ref{sec:phys}. Then, we provide details about the simulations we run with Athena++ in Section \ref{sec:sim}, including the cosmic-ray transport and heating/cooling physics we implemented. In Section \ref{sec:results} we present our results before discussing them in the context of previous work in Section \ref{sec:disc}. Finally, we summarize the outcomes of this work in Section \ref{sec:conc}.

\section{Physical Background} \label{sec:phys}
The $\gamma$-ray luminosity of a plasma with nucleon density $n$ and cosmic ray energy density $E_c$ is \citep{2007Enslin,2008MNRAS.384..251G}:
\begin{equation}
    L_\gamma \propto \int_V d^3x \, \eta_\pi \propto \int_V d^3x \,  n E_c 
    \label{eqn:luminosity}
\end{equation}
where $\eta_\pi$ is the energy lost per unit volume, per unit time by cosmic rays due to hadronic interactions and $V$ is the volume of plasma. We stick to a proportionality because the constant which would be necessary is set by many assumptions about the cosmic ray spectrum \citep{2007Enslin,2008MNRAS.384..251G}. However, regardless of the microphysics, the dependence of $L_\gamma$ on $n$ and $E_c$ is the same.

Assuming a constant (integrated over the cosmic-ray energy spectrum) $L_0$ which captures those assumptions, the total $\gamma$-ray luminosity from that volume $V$ of gas is 
\begin{multline}
    L_\gamma = \frac{L_0}{V} \int_V d^3 x \left(\frac{E_c}{\mathrm{eV}\, \mathrm{cm}^{-3}} \right)
    \left(\frac{n}{\mathrm{cm}^{-3}} \right) \\
    = L_0 \left\langle  \left(\frac{E_c}{\mathrm{eV}\, \mathrm{cm}^{-3}} \right)
    \left(\frac{n}{\mathrm{cm}^{-3}} \right) \right\rangle 
    \label{eqn:luminosity_equals}
\end{multline}
where we have used angled brackets to denote volume averaging. From Equation \ref{eqn:luminosity_equals}, it becomes easier to analyze the $\gamma$-ray emission through a statistics lens: the emission is determined by the correlation of two random variables. After Equation \ref{eqn:luminosity_equals}, we proceed with $E_c$ and $n$ in units of $\mathrm{eV}\, \mathrm{cm}^{-3}$ and $ \mathrm{cm}^{-3}$.

If the density and cosmic ray energy density are two independent random variables, then the luminosity is trivial to calculate:
\begin{equation}
     L_\gamma \rightarrow L_0 \left\langle E_c \right\rangle
    \left\langle n \right\rangle.
    \label{eqn:simple}
\end{equation}
However, we know that cosmic ray energy and electron density are not independent variables. They are coupled together by a variety of physical processes.

Depending on the cosmic ray transport, magnetic field structure, and dynamics of the diffuse gas, the correlation varies significantly. For the sake of simplicity, we define a normalized expectation of the product of cosmic ray energy density and electron density:
\begin{equation}
    \mathcal{C} = \frac{\left\langle  E_c n \right\rangle}{ \left\langle E_c \right\rangle
    \left\langle n \right\rangle}.
    \label{eqn:correlation_def}
\end{equation}
Then the total luminosity is just 
\begin{equation}
    \frac{L_\gamma}{L_0}
    = \left\langle E_c\right\rangle \left\langle n \right\rangle \mathcal{C}.
    \label{eqn:normalized_Lum}
\end{equation}

The reason we formulate the luminosity in the way of Equation \ref{eqn:normalized_Lum} is because it isolates the three factors in determining the $\gamma$-ray luminosity. First, the average cosmic ray energy density $\langle E_c \rangle$. Second, one could change the average density $\langle n \rangle$. Third, one could adjust the correlation of thermal plasma and cosmic ray energy density, which is captured in the term $\mathcal{C}$. 

For a large patch of diffuse gas, the number density $\langle n \rangle$ will change very little unless a runaway cooling process drives the system to form dense, neutral gas. Even then, the integrated number density of particles in a closed system is constant because it is proportional to the total mass. Instead, for a patch of diffuse gas, the other two factors, $\langle E_c \rangle$ and $\mathcal{C}$, can vary significantly as a function of the cosmic ray transport and turbulent properties of the diffuse gas. 

\subsection{Evolution of Average Cosmic Ray Energy Density} \label{sec:phys:cr_energy}

First, we discuss how the average cosmic ray energy density $\langle E_c \rangle$  can change. There could be actual cosmic ray sources (e.g., shocks) which accelerate thermal ions, turning them into cosmic rays. Cosmic ray losses, such as those due to hadronic interactions or coulomb interactions, can reduce the average cosmic ray energy density. Finally, turbulent reacceleration could cause cosmic-ray energy to increase \citep{1988SvAL...14..255P}.

Discrete sources such as shocks (from supernovae or jets) are a significant factor in populating the diffuse gas of the CGM and ICM with cosmic rays. However, in this work, we focus on the evolution of a chunk of diffuse gas which is not actively being shocked or injected with cosmic rays. Our simulations start with a set amount of cosmic-ray energy, and that energy would have originally come from some accelerator. But we do not include the sources themselves, other than in the form of an initial cosmic ray background. This treatment allows us to more easily track the interaction of the background cosmic rays with the turbulent cascade, agnostic to the sources and rate of cosmic-ray production from the many possible accelerators. 

Hadronic interactions and Coulomb interactions are not dominant enough in diffuse gas to completely erase the cosmic ray background. The loss timescale due to hadronic interactions is a fraction of a $\mathrm{Gyr}$, so we include the loss of energy to hadronic interactions. But as we will see, this loss still plays a subdominant role in our simulations.

\begin{figure}
    \centering
    \includegraphics[width=\linewidth]{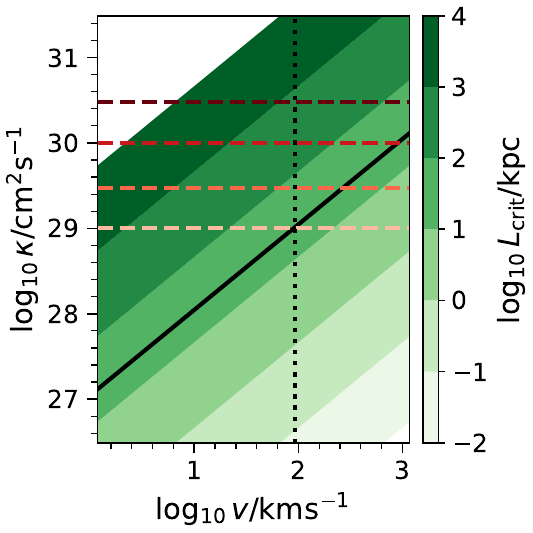}
    \caption{Contours of the critical outer length scale $L_\mathrm{crit}$, where cosmic-ray reacceleration by long-wavelength turbulence is optimal. Green contours show the dependence of this length scale on the cosmic ray diffusion coefficient $\kappa$ and the phase velocity of the turbulence $v$. The solid black line shows the outer scale of turbulent driving used in this work $(24\mathrm{kpc})$, and the dotted black line shows the sound speed of the initial homogeneous medium. The dashed lines correspond to each of the diffusion cases we examine (a different hue of red for each diffusion coefficient we examine). The smallest diffusion coefficient we use, $10^{29}\, \mathrm{cm}^{2}\mathrm{s}^{-1}$, is at the optimal value for our simulation setup.}
    \label{fig:turb_length}
\end{figure}

Finally, turbulent reacceleration is an avenue by which cosmic-ray energy density can increase in any chunk of diffuse gas. If the cosmic rays probe turbulent eddies on a large scale, but decouple from smaller scale eddies, then the cosmic rays can sap energy from the turbulent cascade \citep{1988SvAL...14..255P}. The process derives from the different length scale dependence of the eddy turnover time $(\tau\sim L/v_\mathrm{ph})$ and the cosmic-ray diffusion timescale $(\tau_c \sim L^2 / \kappa)$. The acceleration is optimized when the outer eddy scale is at  a critical length scale $L_\mathrm{crit}$. This critical length is related to the diffusion coefficient of cosmic rays $\kappa$ and the turbulent phase velocity $(v_\mathrm{ph} = \sqrt{(P_g+ P_c B^2/8\pi)/\rho})$ by \citep{2022ApJ...941...65B}:
\begin{equation}
    \kappa \approx 0.15 L_\mathrm{crit} v_\mathrm{ph}.
    \label{eqn:kappa_critical}
\end{equation}

In Figure \ref{fig:turb_length}, we show a contour map of critical length scale $L_\mathrm{crit} $ as a function of diffusion coefficient and turbulent phase velocity. We also plot the outer turbulent driving scale $(24\,\mathrm{kpc})$ we use in our simulations as a solid black line. Each diffusion coefficient we consider is shown as a dashed red line (with different intensities). Finally, we also plot a dotted black line at the turbulent phase velocity in our simulations. We include Figure \ref{fig:turb_length} to make the criteria necessary for turbulent reacceleration clearer. In some of our simulations, we include streaming transport and spatially variable cosmic-ray transport --- both of these processes will lead to a deviation from Equation \ref{eqn:kappa_critical} \citep{2023ApJ...955...64B,2024ApJ...974...17H}.

Figure \ref{fig:turb_length} highlights a key aspect of turbulent acceleration: it primarily happens on large scales $>\mathrm{kpc}$. This requirement is set by the cosmic-ray diffusion coefficient being a relatively large quantity. For example, observations of the local interstellar medium suggest a value of $10^{28}\,\mathrm{cm}^2\,\mathrm{s}^{-1}$ \citep{2019PhRvD..99j3023E}. With a reasonable ISM turbulent phase velocity of $\sim 10\, \mathrm{km}\, \mathrm{s}^{-1}$, the critical length scale will be $> 10\, \mathrm{kpc}$, significantly larger than typical ISM scales. Instead, a much lower diffusion coefficient is necessary for turbulent acceleration to operate efficiently \citep{2024ApJ...974...17H}. 

Only on large scales $(>\mathrm{kpc})$ and in diffuse gas (with a larger $v_\mathrm{ph}$) could turbulent acceleration operate efficiently. In fact, we can estimate the growth rate of cosmic ray energy as a result of this cosmic-ray acceleration via turbulence. Following the work of \cite{2022ApJ...941...65B,2023ApJ...955...64B}, the growth time for anisotropic, magnetic field aligned transport is 
\begin{equation}
    \frac{E_c}{\dot{E}_c} \sim \frac{p^2}{D_{pp}}= t_\mathrm{grow}  \sim 
    \frac{9}{2}\frac{\kappa_\parallel}{v^2} \left( \frac{v L_0}{\kappa_\parallel}\right)^{1/2}.
    \label{eqn:growth_time}
\end{equation}
We use the notation from \cite{2022ApJ...941...65B}: $p$ is cosmic-ray momentum, $D_{pp}$ is the momentum diffusion coefficient, $v$ is the root-mean-square velocity, $\kappa_\parallel$ is the diffusion coefficient parallel to the mean magnetic field, and $L_0$ is the outer scale of the turbulent cascade. \cite{2022ApJ...941...65B} derive this equation by augmenting the usual acceleration time for a large diffusion coefficient \citep{1988SvAL...14..255P} with the number of eddies a cosmic ray interacts with, causing the growth time to decrease $t_\mathrm{grow} \rightarrow t_\mathrm{grow}N^{-1}$. They estimate this number is $N_\mathrm{eddy} \sim \sqrt{\kappa_\parallel / (vL_0)}$, which is then included in the acceleration timescale (Equation \ref{eqn:growth_time}). 

We will see that this formula fails to accurately account for how the cosmic rays interact with multiphase turbulence. In the multiphase case, it is important to remember there is nothing magical about this turbulent acceleration. The acceleration comes directly from the compression of gas, which accelerates cosmic rays through the same mechanism as first-order Fermi acceleration. In a sense, when the cosmic rays can experience more compressions than rarefactions (i.e., they are couple above some length scale and decoupled from smaller scales), then they can get extra energy. However, in multiphase astrophysical turbulence, the compressions are stronger as a result of radiative cooling. The cooling time is generally shorter than the diffusion time, so the cosmic rays are coupled to the cooling gas. But each cold cloud will exist for a significantly longer time than an eddy turn over time if the heating processes are too slow. Therefore, cosmic rays compressed in a cold cloud would be able to leak out of the cold cloud, escaping with the energy they gained during the initial cooling process.

\subsection{Correlation} \label{sec:phys:corr}

\begin{figure}
    \centering
    \includegraphics[width=\linewidth]{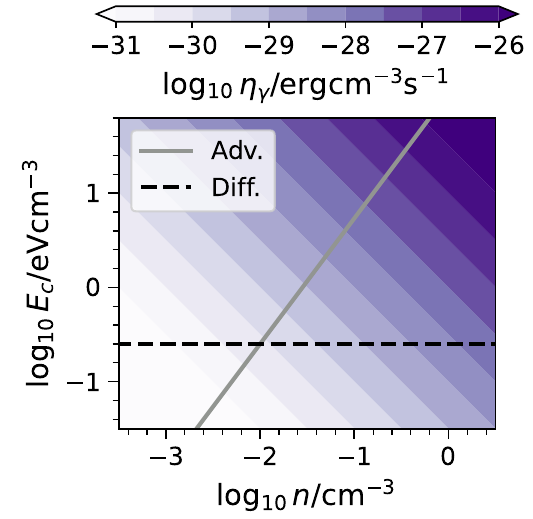}
    \caption{$\gamma$-ray emissivity as a function of cosmic-ray energy density $E_c$ and plasma density $n$. The emissivity is proportional to the product of these quantities. The transport of cosmic rays determines the polytropic relationship between $E_c$ and $n$. Advective transport, when cosmic rays are locked to the gas, leads to a $n^{4/3}$ dependence, illustrated with a solid gray line. More diffusive, decoupled cosmic ray transport leads to a $n^{0}$ dependence which is shown as a dashed black line. Streaming along magnetic flux tubes would produce a $n^{4/3}$ dependence.  As gas compresses and density increases, different cosmic-ray transport mechanisms produce significantly different $\gamma$-ray emissivity. Specifically, more advective (slower) transport leads to more $\gamma$-ray production.}
    \label{fig:Lgam_diagram}
\end{figure}

The correlation factor $\mathcal{C}$ defined in Equation \ref{eqn:correlation_def} has not yet been significantly examined for a turbulent medium. The only previous consideration is in \cite{2010Pinzke}, where they account for the radial variation of the cosmic-ray pressure to gas pressure ratio $X_c = P_c / P_g$, which combined with a density profile could be translated into the correlation factor $\mathcal{C}$ as a function of radius in a galaxy or galaxy cluster. 

In general, $\mathcal{C}$ will be determined by the system's distribution in $E_c$-$n$ phase space, illustrated in Figure \ref{fig:Lgam_diagram}. Figure \ref{fig:Lgam_diagram} shows how the $\gamma$-ray emissivity changes in $E_c$-$n$ phase space. The $\gamma$-ray emissivity is proportional to the hadronic loss rate $\eta_\pi$ in Equation \ref{eqn:luminosity}. The full formula for $\eta_\gamma$ is \citep{2008MNRAS.384..251G}:
\begin{multline}
    \eta_\gamma = \frac{\eta_\pi}{3} =  1.95 \times 10^{-16} \frac{\mathrm{cm}^3}{\mathrm{s}}
    \times E_c n \\
    = 3.12 \times 10^{-28} \frac{\mathrm{erg}}{\mathrm{cm}^3\mathrm{s}} 
    \left( \frac{E_c}{\mathrm{eV}\,\mathrm{cm}^{-3}}\right) 
    \left( \frac{n}{\mathrm{cm}^{-3}}\right).
    \label{eqn:gamma_ray_emissivity}
\end{multline}
Note our $\eta_\pi$ is equivalent to $\Gamma_h$ in \cite{2008MNRAS.384..251G}.  We assume only one third of hadronic interactions produce neutral pions, which will then decay into $\gamma$-rays. So only the neutral pion channel contributes to the $\gamma$-ray emissivity $\eta_\gamma$. Additionally, Equation \ref{eqn:gamma_ray_emissivity} can be used to calculate $L_0$ from Equation \ref{eqn:luminosity_equals}.

Depending on cosmic ray transport, the distribution of the system will move and adjust in different ways in this space. Specifically, for slow transport, the cosmic rays are tied to the thermal gas and follow a $n^{4/3}$ dependence. But for fast transport (diffusive or streaming), the cosmic rays are able to move through gas faster, leading to a $n^0$ dependence where the cosmic-ray pressure becomes uniform despite density variations. In reality, cosmic-ray transport will depend on the local plasma properties, magnetic field strength, etc. As a result, the distribution in the $E_c$-$n$ phase space can be complex. 

Considering $E_c$-$n$ phase space, we can actually see that the independence assumption $(\mathcal{C} =1)$ is equivalent to saying the distribution is concentrated in a single point. For a distribution function $f(n,E_c)$ such that $\int dn \int dE_c \left[f(n,E_c) \right] = 1$, the average density $\langle n \rangle$ and $\langle E_c \rangle$ are ``centers of mass'' in phase space. Similarly, the ``product of inertia''\footnote{The products of inertia are the off-diagonal components of the moment of inertia tensor. See \url{https://en.wikipedia.org/wiki/Moment_of_inertia\#Definition_2}.} of the distribution function would be $I_{nE} = \langle (n-\langle n \rangle) (E_c-\langle E_c \rangle) \rangle$. Thinking in terms of moments of inertia and mass distributions, we can relate the correlation factor to the product of inertia by: 
\begin{equation}
    \mathcal{C} = 1 + \frac{I}{\langle n \rangle \langle E_c \rangle}.
\end{equation}
For an arbitrary distribution in $n$-$E_c$ phase space, the correlation could be much different than $1$. More compact, centered distributions will have $\mathcal{C}\sim 1$. But the distributions we find in our simulations are not simple and compact. 

A final important facet of $E_c$-$n$ phase space is how $\gamma$-ray luminosity heavily favors dense gas with high cosmic ray energy density. These regions will produce significantly more $\gamma$-rays. So, to avoid any small regions of denser gas dominating the $\gamma$-ray luminosity, the cosmic rays must de-correlate from the dense gas.

De-correlation means that cosmic ray transport must be fast, but most importantly it must be fast in dense gas. We will increase the speed of cosmic-ray transport from the estimate of the Milky Way diffusion coefficient of $\kappa_\parallel \sim 10^{28}\mathrm{cm}^{2}\mathrm{s}^{-1}$ in two ways. First, we will simply increase the diffusion coefficient. Second, we will examine streaming transport (the self-confinement model), when the Alfv\'en speed $B/\sqrt{4\pi \rho}$ dominates transport. Finally, in another simulation, we will increase Alfv\'en speed in cold gas to account for the effect of ion-neutral damping.

\section{Simulation Details}\label{sec:sim}
We examine the evolution $\gamma$-ray luminosity in a turbulent box of multiphase gas using Athena++ \citep{2018Jiang,2020Stone}. Athena++ solves the following equations:
\begin{equation}
    \pdv{\rho}{t} + \div{\rho \vb{u}}=0
    \label{eqn:continuity}
\end{equation}
\begin{multline}
    \pdv{\rho \vb{u}}{t} + \div{ \left( \rho \vb{u} \vb{u} + \mathbb{1}\left(P_g + \frac{B^2}{2}\right) - \vb{B}\vb{B}\right)} \\
    = \sigma_c \vdot \left( \vb{F}_c - \gamma_c \vb{u} E_c \right)
    \label{eqn:momentum}
\end{multline}
\begin{equation}
    \pdv{\vb{B}}{t} = \curl{\left( \vb{u} \times \vb{B}\right)}
    \label{eqn:induction}
\end{equation}
\begin{multline}
    \pdv{E_\mathrm{MHD}}{t} + \div{\left( \vb{u}\left(E_\mathrm{MHD} + P_g + \frac{B^2}{2} \right)  - \vb{B}\left(\vb{B}\vdot \vb{u}\right)\right)} \\
    =  \left(\vb{u} + \vb{v}_s \right) \vdot \sigma_c \vdot \left( \vb{F}_c - \gamma_c \vb{u} E_c ) \right) - \rho \mathcal{L}
    \label{eqn:energy}
\end{multline}
\begin{multline}
    \pdv{E_c}{t} + \div{\vb{F}_c} = \\ 
    - \left(\vb{u} + \vb{v}_s \right) \vdot \sigma_c \vdot \left( \vb{F}_c - \gamma_c \vb{u} E_c ) \right) 
    - \frac{E_c}{\tau_h}
    \label{eqn:cr_energy}
\end{multline}
\begin{equation}
    \frac{1}{V_m^2}\pdv{\vb{F}_c}{t} + (\gamma_c - 1)\grad{E_c}  
    = - \sigma_c \vdot \left( \vb{F}_c - \gamma_c \vb{u} E_c \right).
    \label{eqn:cr_flux}
\end{equation}
In the above, $\rho$ is the plasma density, $\vb{u}$ is the bulk velocity of the plasma, $P_g$ is the gas pressure, $\vb{B}$ is the magnetic field (with strength $B$), $\sigma_c$ is the cosmic-ray transport matrix (see Equation \ref{eqn:cr_transport_matrix} in Section \ref{sec:sim:cr_phys}), $\vb{F}_c$ is the cosmic-ray energy flux density, $\gamma_c = 4/3$ is the adiabatic index of the cosmic-ray fluid, $\tau_h$ is the loss-rate of cosmic-ray energy density via hadronic interactions (see Equation \ref{eqn:hadronic_loss_time} in Section \ref{sec:sim:cr_phys}), $E_\mathrm{MHD} = \rho u^2 /2 + B^2/2 + P_g/(\gamma_g - 1)$ is the total magnetohydrodynamic energy, $\rho \mathcal{L}$ is the heat-loss function (see Equation \ref{eqn:heat_loss} in Section \ref{sec:sim:cool_heat}), $\vb{v}_s$ is the cosmic-ray streaming velocity (see Equation \ref{eqn:streaming_vel} in Section \ref{sec:sim:cr_phys}), and $V_m$ is the modified speed of light. 

The computational methods we use within Athena++ are relatively standard. However, we still specify the non-thermal physics related to cosmic rays and optically thin radiative cooling and heating for completeness.

\begin{deluxetable}{lccc}
    \tablewidth{\linewidth}
    \tablecolumns{4}
    \tablecaption{Simulation Parameters}
    \label{tab:sims}
    \tablehead{\colhead{Simulation} & \colhead{$\kappa_\parallel / \mathrm{cm}^{2}\mathrm{s}^{-1}$} & \colhead{Streaming?} & \colhead{IND?} }
    \startdata
    \texttt{1e29} & $1\cdot 10^{29}$ & No & No \\
    \texttt{3e29} & $3\cdot 10^{29}$ & No & No \\
    \texttt{1e30} & $1\cdot 10^{30}$ & No & No \\
    \texttt{3e30} & $3\cdot 10^{30}$ & No & No \\
    \texttt{Str} & $3\cdot 10^{28}$ & Yes & No \\
    \texttt{StrIND} & $3\cdot 10^{28}$ & Yes & Yes \\
    \enddata
    \tablecomments{IND is short for ion-neutral damping. We include this effect through a parameterization of the ionization fraction $f_i$ in Equation \ref{eqn:ionization_fraction}.}
\end{deluxetable}

\subsection{Setup}\label{sec:sim:setup}

Our simulation's initial setup is a homogeneous box with side lengths of $48\,\mathrm{kpc}$, filled with diffuse ionized plasma, a uniform magnetic field, and a background of cosmic rays. We stir this box, using an Ornstein-Uhlenbeck process, and force the velocity perturbations to all together have a combined kinetic energy injection rate of $\dot{E}_\mathrm{inj} = 1.5 \times 10^{41} \mathrm{erg}\,\mathrm{s}^{-1} $. This low level of energy injection is enough to drive the sub-sonic turbulence in the diffuse plasma. We want the simulation to have low Mach numbers because that matches observations of the ICM and CGM \citep{1986Sarazin}.

The initial parameters for each simulation are $T_0 = 10^7\mathrm{K}$, $\rho_0 =10^{-2} m_H\mathrm{cm}^{-3}$, $\beta_0 = 8 \pi P_g / B^2 = 50$, and $X_\mathrm{cr} = P_c/P_g = 0.01$. These correspond to a magnetic field strength of $B= 2.6\mu\mathrm{G}$ and a cosmic-ray pressure of $P_\mathrm{cr } = 8.6\times 10^{-2}\mathrm{eV}\,\mathrm{cm}^{-3}$. In Table \ref{tab:sims}, we show the cosmic-ray transport parameters for each simulation.

Also, given we now have a set box size and initial conditions, we can define $L_0$ from Equation \ref{eqn:luminosity_equals} by multiplying Equation \ref{eqn:gamma_ray_emissivity} by the simulation volume:
\begin{equation}
    L_0 
    = 2.63 \times 10^{39} \frac{\mathrm{erg}}{\mathrm{s}}
    X_\mathrm{cr}  \left(\frac{n}{10^{-2}\mathrm{cm}^{-3}} \right)^2 
    \left( \frac{T}{10^7 \mathrm{K}}\right).
    \label{eqn:luminosity_0_estimate}
\end{equation}

The luminosity in Equation \ref{eqn:luminosity_0_estimate} corresponds to a flux of $\sim 1\mathrm{GeV}$ photons of $\sim 1.7 \times 10^{-12}\mathrm{cm}^{-2}\mathrm{s}^{-1}$ assuming the simulation is $100\mathrm{Mpc}$ away from the detector. That estimate is more useful for comparison with galaxy cluster observations \citep{2014Ackermann}, even though our simulation size is significantly smaller than a cluster. In fact, for studying an entire cluster, we would need to decrease our initial density to $10^{-4}\mathrm{cm}^{-3}$. The density we chose allows us to begin to study diffuse gas in general, as opposed to the ISM.

\subsection{Cosmic-Ray Physics}\label{sec:sim:cr_phys}

We use the two-moment method implemented in \cite{2018Jiang}. This implementation allows us to include the impacts of multiple cosmic-ray transport processes. 

First, it includes the advection of cosmic rays with the background plasma, at the bulk flow velocity $\vb{u}$ in Equations \ref{eqn:continuity}-\ref{eqn:cr_flux}. Advection derives from the fact that cosmic rays are scattering off magnetic field fluctuations. As the cosmic rays scatter in pitch angle, the cosmic-ray distribution function evolves in the frame co-moving with the waves, which includes any bulk flow speed \citep{2017PhPl...24e5402Z}. 

Secondly, the wave frame also includes the streaming speed $\vb{v}_s$. The exact value of this speed depends on the source of the magnetic fluctuation which scatter the cosmic rays \citep{2017PhPl...24e5402Z}. In the  self confinement model, the fluctuations are Alfv\'en waves which are  generated by the cosmic rays themselves as they stream down their pressure gradient, in which case $\vb{v}_s= v_A$ and is directed down $\grad{P_c}$. In the  extrinsic turbulence model, the fluctuations are generated by a turbulent cascade which is generally taken to be balanced; i.e., the waves propagate up and down $\grad{P_c}$ with equal intensity, and $\bv{v}_s\equiv 0$. More general constructions allow for a continuous range between the self-confinement and extrinsic turbulence models, but in this setup we switch between the two models discontinuously. In Table \ref{tab:sims}, the simulations with ``no'' streaming have $\vb{v}_s = 0$, which corresponds to the extrinsic turbulence model. In simulations with streaming (``Yes'' in third column of Table \ref{tab:sims}), we follow the self-confinement model and include the streaming speed as 
\begin{equation}
    \vb{v}_s = -\frac{\vb{B}}{\sqrt{4\pi \rho_i}} \frac{\vb{B} \cdot \grad{P_c}}{\left|\vb{B} \cdot \grad{P_c} \right|},
    \label{eqn:streaming_vel}
\end{equation}
where $\rho_i$ is the ion density. Except in the simulation with ion-neutral damping (\texttt{StrIND}), we assume the plasma is fully ionized regardless of temperature $(\rho_i = \rho)$.

Finally, we also include the spatial diffusion, which derives from the pitch angle scattering of cosmic rays. This diffusion appears regardless of the source of the scattering magnetic field fluctuations \citep{2017PhPl...24e5402Z}. Assuming the mean magnetic field is resolved, the transport will be anisotropic, with cosmic rays diffusing faster in the direction of the mean magnetic field than in the directions perpendicular to it. To include this effect, we adjust the combined transport matrix $\sigma_c$ in Equations \ref{eqn:continuity}-\ref{eqn:cr_flux} to be
\begin{equation}
    \sigma_c = \frac{1}{\kappa_\perp} \mathbb{1} + \left( \frac{1}{\kappa_\parallel - \frac{4 E_c \vb{v}_s \cdot \hat{b}}{\hat{b}\cdot \grad E_c}} - \frac{1}{\kappa_\perp}\right) \hat{b}\hat{b}
    \label{eqn:cr_transport_matrix}
\end{equation}
where $\hat{b} = \vb{B}/\left|\vb{B } \right|$ is the direction of the magnetic field, $\kappa_\perp$ is the cosmic-ray diffusion coefficient perpendicular to the magnetic field, and $\kappa_\parallel$ is the cosmic-ray diffusion coefficient parallel to the magnetic field. 

An approximate way of including the effect of ion-neutral damping (IND in Table \ref{tab:sims}) is to adjust the streaming velocity to account for the appearance of neutrals in thermal gas with a temperature below a certain value. Only in the \texttt{StrIND} simulation, we model the ion density as 
\begin{equation}
    \rho_i = \rho f_i(T) 
\end{equation}
where $\rho$ is the local gas density and $f_i(T)$ is the ion fraction at a given temperature. We model the ion fraction with a simple switching function $S(x)$:
\begin{equation}
    f_i(T) = 1 + (f_\mathrm{min} - 1)  S\left( \frac{T-T_i}{\Delta T}\right).
    \label{eqn:ionization_fraction}
\end{equation}
We choose $S(x) = 0.5 \left[ 1+ \tanh(x) \right]$, $T_i = 10^5\mathrm{K}$, and $\Delta T = 10^4\mathrm{K}$. This formulation is similar to the setup others have used to study the effect of changes in ionization. 

For a fully realistic implementation, the temperature of the switch from fully ionized to partially ionized should be lowered to $T_i \sim 10^4\,\mathrm{K}$ \citep{1993ApJS...88..253S,2021ApJ...913..106B}.  However, our cooling curve peaks at $ 10^5\,\mathrm{K}$(see Section \ref{sec:sim:cool_heat}), and that temperature sets the dividing line between hot and cold plasma in our simulation. Choosing the switching temperature this way minimizes the impact of our choice of switching function $S$ and transition width $\Delta T$. Therefore, our setup allows us to isolate the effect of well-defined regions (cold clouds in a diffuse medium) with a different cosmic-ray transport. A more complex setup would require more resolution at cloud interfaces, and the results would be more dependent on our choice of the switching function $S$ and the width of the switch $\Delta T$.

We also include the losses due to hadronic interactions, with a simple energy loss time $\tau_h$ (see Equation \ref{eqn:cr_energy}). Using Equation 12 from \cite{2008MNRAS.384..251G} for the energy lost to hadronic interactions per unit volume $\Gamma_h$, the loss time is:
\begin{multline}
    \frac{1}{\tau_\pi} = \frac{\Gamma_h}{E_c} = 5.86 \times 10^{-16} \mathrm{s}^{-1} \left(\frac{n}{\mathrm{cm}^{-3}} \right) \\
    = 0.185  \, \mathrm{Gyr}^{-1} \left(\frac{n}{ 10^{-2 }\mathrm{cm}^{-3}} \right).
    \label{eqn:hadronic_loss_time}
\end{multline}
The rate $5.86 \times 10^{-16} \mathrm{s}^{-1}$ comes directly from \cite{2008MNRAS.384..251G}. Since the total loss rate is proportional to $n E_c$, the loss time is inversely proportional to just the density $n$. It is also important to note that this rate leads to a loss of energy because of the sign of  $E_c / \tau_\pi$ in Equation \ref{eqn:cr_energy}.

\begin{figure*}
    \centering
    \includegraphics[width=\linewidth]{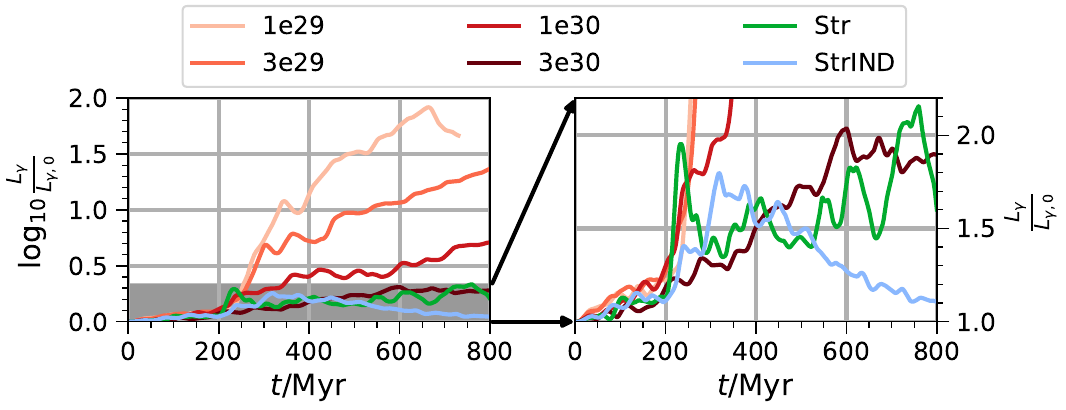}
    \caption{Evolution of $\gamma$-ray luminosity over time for each simulation. Left side shows the evolution of the logarithm of $L_\gamma$ relative to the initial state. Right side shows a zoomed in, linear scale version to highlight the differences between the various fast transport cases. While slow transport (lighter red lines) leads to a nearly $100$ times larger $L_\gamma$ (left plot), there are order unity discrepancies between the different fast transport results (right plot).}
    \label{fig:Lgamma}
\end{figure*}

\subsection{Heating \& Radiative Cooling} \label{sec:sim:cool_heat}

In this section, we detail the heating and cooling in the heat loss function $\rho \mathcal{L}$ \citep{1965Field} in Equation \ref{eqn:energy}. We use the common formulation which assumes a constant heating rate $\Gamma$ and a temperature dependent volumetric cooling function:
\begin{equation}
    \rho \mathcal{L} = n^2 \Lambda(T) - n \Gamma.
    \label{eqn:heat_loss}
\end{equation}
We calculate the heating and cooling terms separately, using an operator splitting approach.

It is possible to connect the evaluation of each term into a single calculation when $\Gamma$ is a constant or a function of temperature \citep{2024MNRAS.527.9683T}. However, it is not so simple when attempting to keep the entire simulation in constant energy balance. Enforcing conservation of energy when including the effects of optically thin radiative cooling requires adjusting $\Gamma$ to counter the average cooling rate throughout the simulation. After evaluating the loss of energy due to the $n^2 \Lambda(T)$ term in Equation \ref{eqn:heat_loss} everywhere, we sum the total energy loss over the entire simulation. Then we subtract off the turbulent energy injection rate. The remaining energy is added back through the $\Gamma$ in Equation \ref{eqn:heat_loss}, a constant rate for every cell in the simulation, conserving total energy. If the turbulent energy injection is larger than the cooling, then we do not add any heating. While this could cause net energy gain, the effect only matters for a short time at the beginning of the simulation. Once large enough temperature variations form as the result of turbulent driving, the regions with more cooling than the average will cool, and regions with less cooling than average will be heated. 

We add the heating to each cell by assuming a constant $\Gamma$. For the calculation of cooling,  we implement the exact integration method from \cite{2009Townsend}. This process is straightforward for power-law cooling curves, so we use a common parameterization of the collisional-ionization equilibrium (CIE) curve from \cite{1995ApJ...440..634R}. Specifically, a piecewise power law given by:
\begin{equation}
    \Lambda(T) = 
    \left\{ 
    \begin{array}{lr}
        0 & T \leq 300 \\
        2.2380 \cdot 10^{-32} T^2 &
        300 \leq T < 2000 \\ 
        1.0012 \cdot 10^{-30} T^{1.5} &
        2000  \leq T < 8000  \\ 
        4.624 \cdot 10^{-36} T^{2.867} &  8000 \leq T < 10^5  \\
        1.78 \cdot 10^{-18} T^{-0.65} &  10^5 \leq T < 4 \cdot 10^7   \\
        3.2217 \cdot 10^{-27} T^{0.5} &  4 \cdot 10^7 \leq T 
    \end{array}
    \right.
    \label{eqn:Cooling}
\end{equation}
where $T$ is in units of $\mathrm{K}$ and $\Lambda$ is in units of $\mathrm{erg}\,\mathrm{cm}^3 \mathrm{s}^{-1}$.

With the heating set to balance the radiative cooling of the thermal, there is only one true loss process in our simulations once turbulence is saturated: the hadronic losses of cosmic ray energy. Since some simulations end up with significantly larger cosmic-ray energy densities on average, they end up with a larger total energy lost to hadronic interactions. This loss causes the total energy in the simulation to decrease over time, but we stop the simulations before lost energy becomes greater than $10\%$ of the initial total energy.

\begin{figure*}
    \centering
    \includegraphics[width=\linewidth]{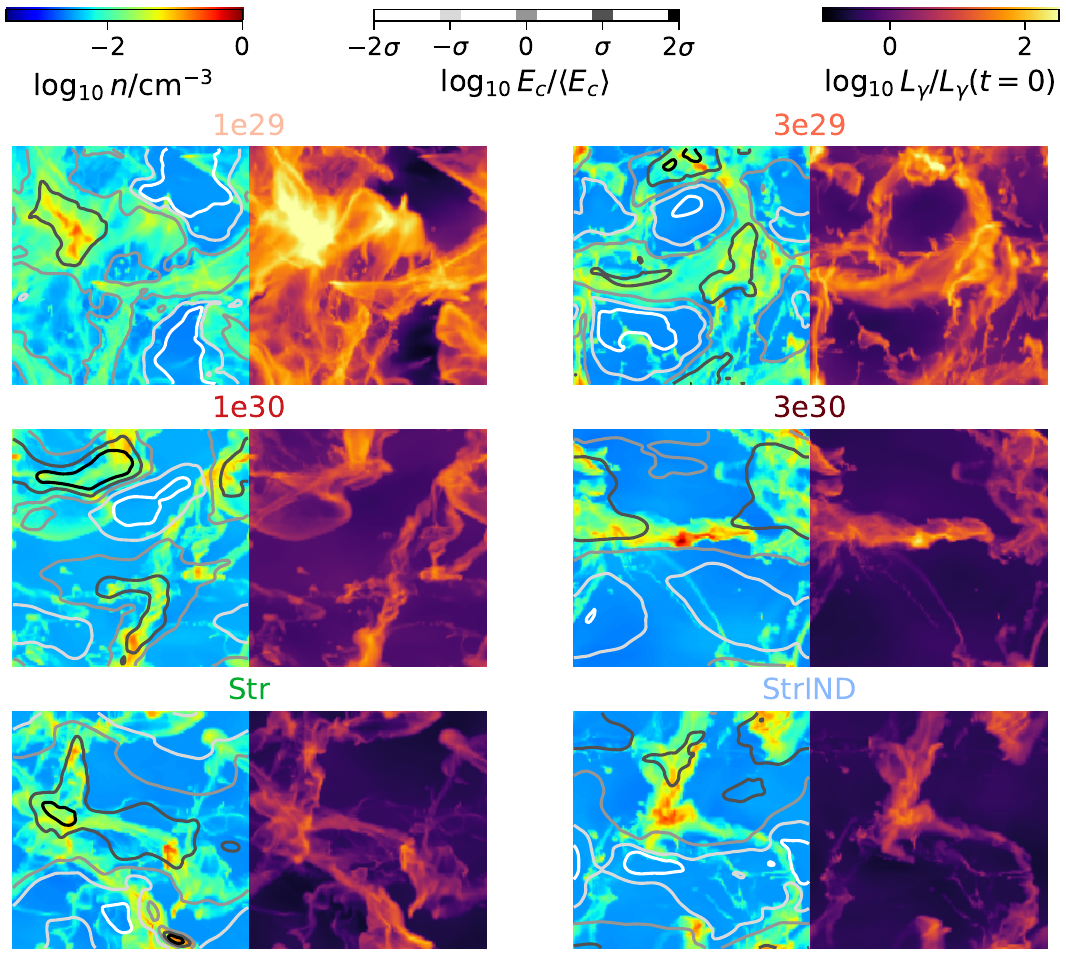}
    \caption{Line-of-sight averaged gas density and cosmic-ray energy density adjacent to integrated $\gamma$-ray luminosity for each simulation. As diffusion coefficient increases, cosmic rays de-correlate from the dense gas and the diffuse $\gamma$-ray luminosity decreases. Note that the cosmic-ray energy density contours are normalized by the total cosmic-ray energy in the simulation to account for the effects of reacceleration, which cause the total luminosity in the \texttt{1e29} run to be 100 times larger than the fast transport runs (see Figure \ref{fig:Lgamma}). For the streaming runs, the $\gamma$-ray luminosity is low, but the cosmic rays are even more de-correlated from dense gas when the effects of ion-neutral damping are included.}
    \label{fig:render}
\end{figure*}

\section{Results} \label{sec:results}

Our primary result is illustrated in Figure \ref{fig:Lgamma}. Each simulation's volume integrated $\gamma$-ray luminosity is shown with respect to the value at $t=0$, which is $L_0=$ for all simulations. The left-side plot in Figure \ref{fig:Lgamma} uses a log-scaled y-axis, whereas the right-side plot zooms in to a linear scale y-axis. The right-side plot clarifies the variation in $\gamma$-ray luminosity across the three fast transport cases (simulations \texttt{3e30}, \texttt{Str}, and \texttt{StrIND} in Table \ref{tab:sims}). The lines with different hues of red are the simulations with only diffusive transport, whereas the green and blue lines correspond to the two simulations with streaming transport and weak diffusion.

Figure \ref{fig:Lgamma} shows that a simple change in cosmic ray transport leads to variations in the total $\gamma$-ray luminosity of 100 times the initial $\gamma$-ray luminosity. Fast transport is necessary for our simulations to be in reasonable agreement with observational upper limits on diffuse emission. However, among the fast transport processes there is variation. Therefore, our simulations additionally illustrate how the microphysics of cosmic ray transport significantly affect observables.

Comparing to our estimate of flux from an object $100 \mathrm{Mpc}$ away from us, the two orders of magnitude increase in luminosity for the \texttt{3e29} simulation is alarming. Propagating through, this simulation (significantly smaller than a galaxy cluster) would have produced a detectable flux $10^{-10}\mathrm{ph}\,\mathrm{cm}^{-2}\,\mathrm{s}^{-1}$ (see Section \ref{sec:sim:setup}).

The only simulation with a decrease in $\gamma$-ray luminosity is the \texttt{StrIND} simulation with increased streaming speed in denser gas. Since this simulation assumes the self-confinement model, cosmic rays are generating their own waves, which then dissipate into the background plasma. This process leads to a decrease in cosmic-ray energy, especially since the heating rate (proportional to the Alfv\'en speed) increases in dense gas. So, once the simulations form enough dense gas, the cosmic rays begin to lose a significant amount of their energy to the background plasma. But as we will see in Section \ref{sec:results:thermo}, that extra heating is marginal from the perspective of the background plasma. 

Figure \ref{fig:render} shows actual snapshots of the simulations at $t=600\mathrm{Myr}$, illustrating how the cosmic-ray energy density is able to de-correlate from the dense gas. This process helps keep the $\gamma$-ray luminosity low in the fast transport simulations. The left plots for each simulation show the average gas density along the (not pictured) $\hat{z}$ axis as a colormap, with contours overlaid highlighting where the average cosmic-ray energy density (again along the $\hat{z}$ axis) is above or below the mean value for the entire simulation. The right plot of each simulation shows the $\gamma$-ray luminosity $L_\gamma$ with respect to the luminosity of the homogeneous initial conditions. 

While the contours of large cosmic-ray pressure track well with dense gas in the \texttt{1e29} simulation, as we move between simulations, we see the cosmic rays and dense gas become more and more de-correlated. The largest de-correlation appears in the \texttt{StrIND} and \texttt{3e30} simulations. The plots of $L_\gamma$ also change significantly as a function of cosmic-ray transport. 

In the rest of this section, we provide more detailed results. We check the thermodynamic state of each simulation to confirm that the only driver for the variation in $\gamma$-ray luminosity is cosmic ray transport. Then, we examine the causes of the increase in $\gamma$-ray luminosity. We find the primary contributor is turbulent acceleration by condensing cold clouds. However, the correlation of cosmic rays and cold gas (defined in Equation \ref{eqn:correlation_def}) also has an impact. Finally, we illustrate how the pockets of cold gas in our simulation set the $\gamma$-ray luminosity, instead of the volume-filling diffuse gas.

\begin{figure*}
    \centering
    \includegraphics[width=\linewidth]{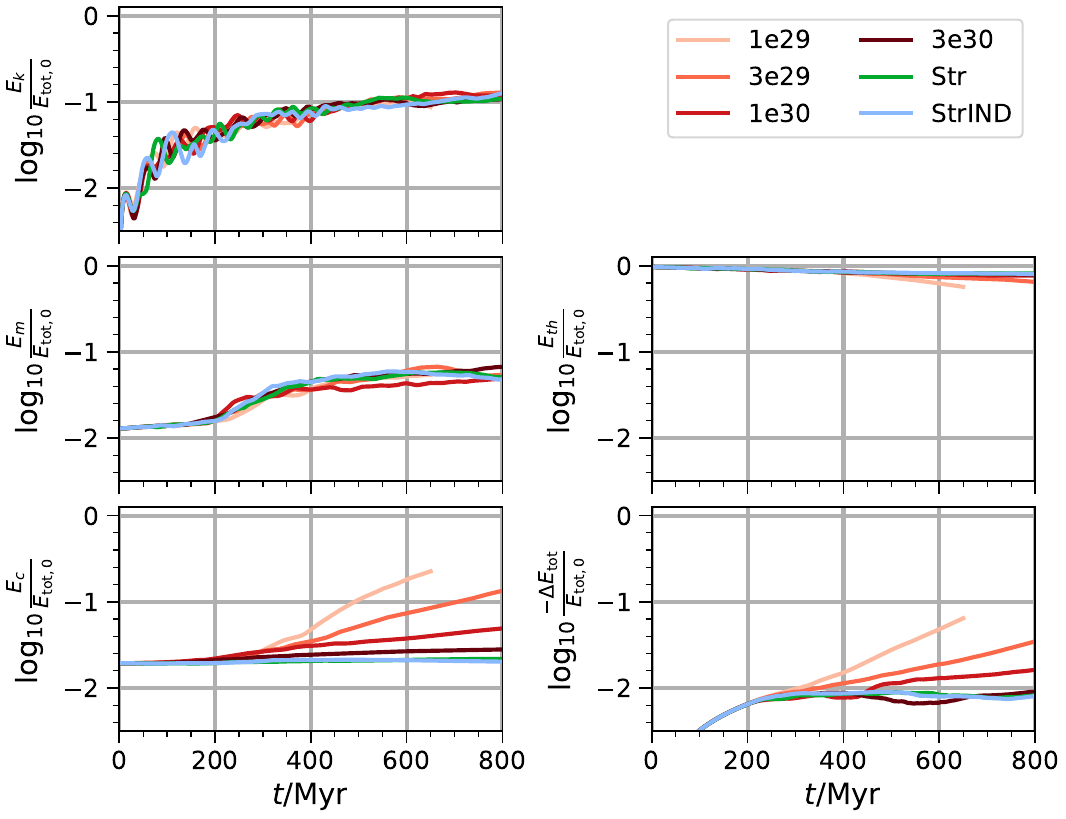}
    \caption{Evolution of total kinetic energy (upper left), total magnetic energy (middle left), total cosmic-ray energy (lower left), total thermal energy (middle right), and total energy change (lower right) for each simulation. The simulations have similar kinetic and magnetic energy evolution. The cosmic-ray energy evolution differs across the simulations because of the different cosmic ray transport in each simulation. Similarly, the thermal energy barely changes, except for in the slow cosmic ray transport cases. This change is clearer in the total energy loss plot (bottom right). The total energy decreases because of $\gamma$-ray emission, which increases proportionally with the cosmic-ray energy density. So, when the cosmic-ray energy density increases (bottom left plot), the total energy loss also increases.  }
    \label{fig:energies}
\end{figure*}

\begin{figure*}
    \centering
    \includegraphics[width=\linewidth]{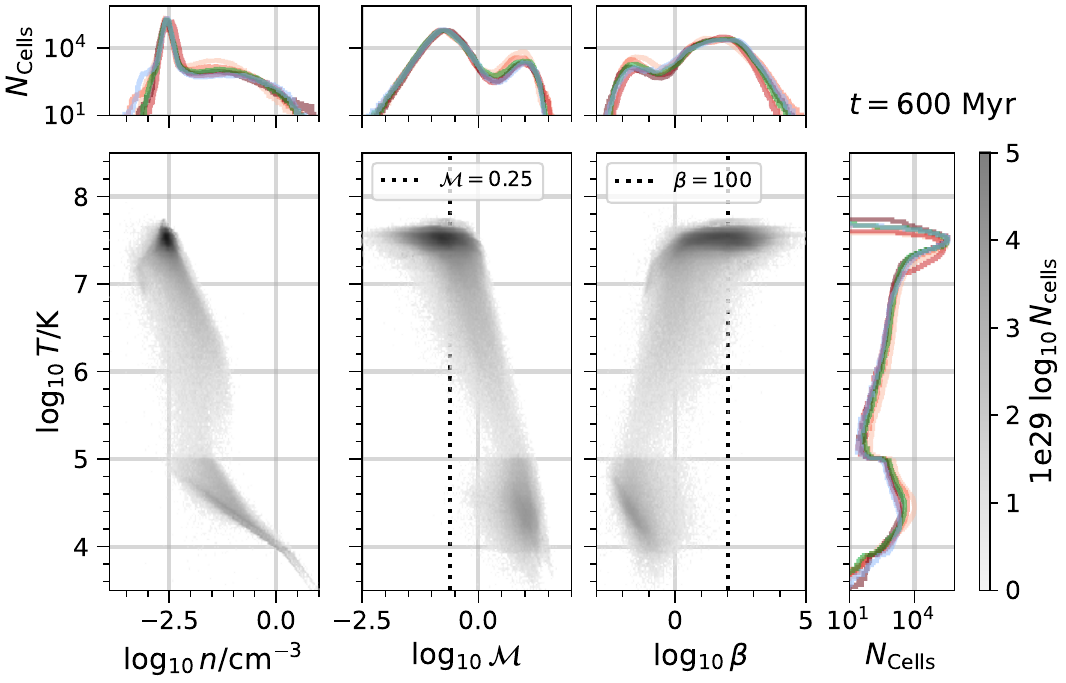}
    \caption{Thermodynamic, kinematic, and magnetic properties of the simulations at $t=600\,\mathrm{Myr}$. Left column shows $T$-$\rho$ phase space, middle column shows $T$-$\mathcal{M}$ phase space, and the right column shows $T$-$\beta$ phase space. In the top row and right of the third column, we show the marginal distributions of gas density $\rho$, Mach number $\mathcal{M}$, plasma beta $\beta$, and temperature $T$. Except for some small deviations in amplitude, the marginal distributions and overall phase space distributions are similar for all simulations.}
    \label{fig:thermodynamics}
\end{figure*}

\subsection{Different $\gamma$-ray Luminosity, Similar Thermodynamics}\label{sec:results:thermo}

As illustrated in Figure \ref{fig:Lgamma}, the simulations have significantly different $\gamma$-ray luminosities. However, they are similar in most other aspects. In Figure \ref{fig:energies}, we show the evolution of each energy component in the simulations. Total kinetic energy and total magnetic energy are consistent across the simulations (left column, top and middle graphs). However, due to turbulent acceleration of cosmic-rays, the total cosmic ray energy grows in some simulations (left column, bottom plot). This energy gain comes at the expense of thermal energy (right column, middle plot). 

Finally, in those simulations where cosmic-ray acceleration occurs, we have a larger change in the energy accounting (right column, bottom plot). Our heating only accounts for the cooling terms and turbulent driving, making sure that the net energy lost via cooling is injected back into the simulation in either kinetic or thermal energy. However, we also included the loss of cosmic-ray energy due to hadronic interactions. This energy loss is proportional to the total cosmic-ray energy, so the hadronic losses increase with time in those simulations where cosmic-ray acceleration occurs. Since we do not add the hadronic losses back in as heating, the total energy of the simulations decreases (see Sections \ref{sec:sim:cr_phys} and \ref{sec:sim:cool_heat}). The initial change in energy is the result of turbulence not being saturated (see the evolution of kinetic energy in the left column, top plot). During this time, the heating is weakened because the accounting assumes the turbulent cascade is in a steady state. This saturation time delay leads to a loss of energy of order $1\%$.

Next, we examine the thermodynamic and turbulent phase spaces for the simulations. In Figure \ref{fig:thermodynamics} we show the simulations mostly overlap in $T$-$\rho$ phase space (bottom row, left plot), $T$-$\mathcal{M}$ (Mach number) phase space (bottom row, middle-left plot), and $T$-$\beta$ phase space (bottom row, middle-right plot). We also show the marginal distributions of temperature $T$ (right-most plot), density $\rho$ (top row, left plot), Mach number $\mathcal{M}$ (top row, middle plot), and beta $\beta$ (top row, right plot).

In the 2D-phase spaces, each simulation is plotted with a black colormap (shown in the colorbar on the right of Figure \ref{fig:thermodynamics}). The histogram of each simulation has the same colormap and shading. The simulations almost completely overlap in each phase space. This similarity in thermodynamic, turbulent, and magnetic state is also apparent in the marginal distributions. The biggest differences are in the Mach numbers of the cold gas in the \texttt{1e29} simulation, and some additional heating in diffuse gas in the \texttt{StrIND} simulation. Similarly, the most common temperature values vary slightly in the marginal temperature distributions. This extra cooling is present primarily because of the cosmic-ray acceleration and subsequent loss of cosmic-ray energy to hadronic interactions in the diffusion simulations (see total energy error plot in Figure \ref{fig:energies}).

Otherwise, the phase space structures in Figure \ref{fig:thermodynamics} are predictable. Rapid cooling leads to a clump of cells between $T\sim 10^4\mathrm{K}$ and $T\sim 10^{5} \mathrm{K}$. Those cells have a larger Mach number, since their sound speed is lower, but the turbulent velocities are similar to the surrounding medium. Those colder, denser cells also have a lower $\beta$, as the magnetic field becomes significantly compressed and strengthened. Additionally, the magnetic field in the dense clouds remains connected to the diffuse medium, so cosmic rays can move between the two phases along field lines as described in Section \ref{sec:sim:cr_phys}.

Overall, the conclusion is that adjusting the cosmic-ray transport will primarily impact the $\gamma$-ray luminosity of the diffuse plasma, not the thermodynamic properties. Any changes to the thermodynamic, turbulent, and magnetic properties of the plasma are minimal. As we discussed in Section \ref{sec:phys:cr_energy}, the primary way for cosmic rays to influence the thermodynamics and turbulence is via turbulent acceleration, but even this has a small impact because the total loss to hadronic interactions due to the increasing cosmic-ray energy density is small compared to the total plasma energy. Again, we are in the limit of $X_{c} \lesssim 1$, which is necessary for the diffuse $\gamma$-ray emission to agree with observed upper limits \citep{2014Ackermann}. 

\begin{figure}
    \centering
    \includegraphics[width=\linewidth]{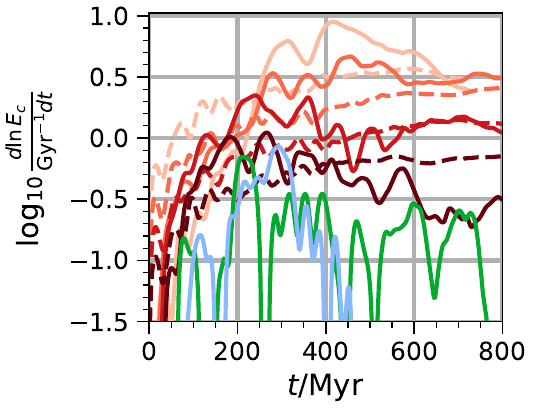}
    \caption{The evolution of $L_\gamma$ is partly a function of the cosmic-ray acceleration timescale. Solid lines show the growth rate (logarithmic time derivative) of the total cosmic-ray energy for each simulation, with line colors matching the legend in Figure \ref{fig:Lgamma}. In dashed lines, we also include the analytic theory expectation for the growth of cosmic-ray energy via turbulent reacceleration \citep{2022ApJ...941...65B}. }
    \label{fig:growth_time}
\end{figure}

\subsection{Turbulent Reacceleration, Cold Cloud Reacceleration, or Correlation? } \label{sec:results:TurbvsCorr}

Now, we quantify the separate impacts of turbulent reacceleration and correlation on the $\gamma$-ray luminosity. We find that when turbulent reacceleration occurs in the slow diffusion simulations, it dominates the $\gamma$-ray luminosity. For fast transport, correlation is the primary factor influencing the $\gamma$-ray luminosity. 

In Figure \ref{fig:growth_time}, we plot the inverse of the growth time of cosmic-ray energy during each simulation. We also plot the inverse growth time from analytic theory for the diffusive simulations (see Equation \ref{eqn:growth_time}), following the work of \cite{2022ApJ...941...65B}. There is significant disagreement in the shape of the simulated curves and the theoretical ones, which derives from the theory not accounting for a multiphase medium. However, ordering of the curves (i.e., the dependence on $\kappa_\parallel$) . At some points, our simulations even get faster acceleration rates than analytic theory predicts. For the streaming simulations, this turbulent acceleration is negligible, and the simulations spend more time with actual negative growth rates (lines dive off the bottom of the logarithmic plot). For reference, we smoothed the growth rates from the simulations with a simple moving average ($0^\mathrm{th}$ order Savitzky-Golay filter) with a window of $50\,\mathrm{Myr}$.

Until now, we have been referring directly to Ptuskin's turbulent reacceleration as the source of the increase in cosmic-ray energy. However, the deviation from analytic theory suggests there could be a different mechanism involved in reaccelerating cosmic rays. To discern the cause of this deviation from theory, we identify where reacceleration occurs in the simulation.

\begin{figure}
    \centering
    \includegraphics[width=\linewidth]{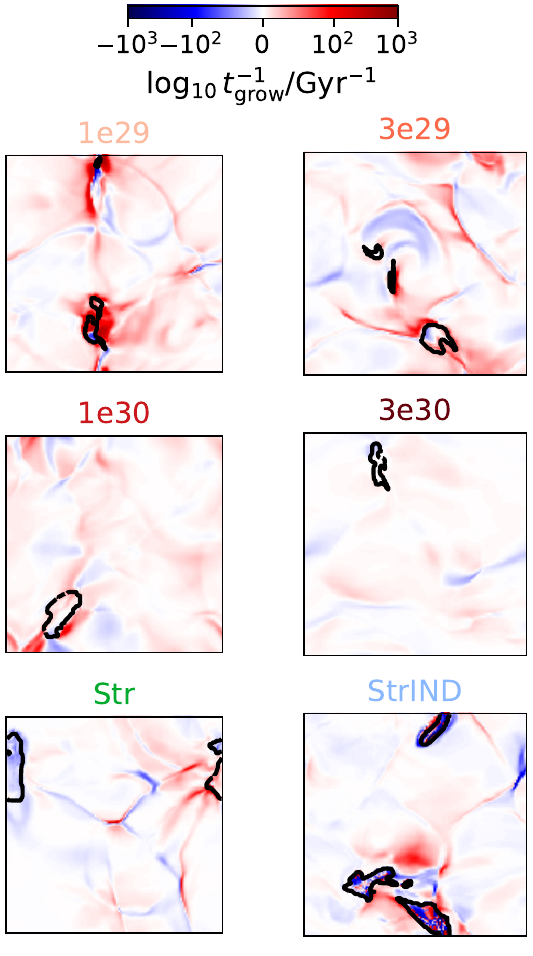}
    \caption{Cosmic-ray energy growth time (negative values are energy loss) in a slice of the simulations at $t=300\,\mathrm{Myr}$. Black contours are where $T=10^5 \,\mathrm{K}$. The growth time calculation ignores hadronic losses, focusing on the interaction with turbulence and the cooling-induced collapse of clouds. The regions on and around the contours are predominantly red, meaning that cooling, condensing clouds preferably increase cosmic-ray energy density. }
    \label{fig:GrowthTimeVisual}
\end{figure}

In Figure \ref{fig:GrowthTimeVisual}, we examine the spatial distribution of cosmic-ray energy growth time at $t=300\,\mathrm{Myr}$ for every simulation. The colormap shows where the cosmic-ray energy is either increasing (red, positive growth rate) or decreasing (blue, negative growth rate) The growth rate is defined as 
\begin{multline}
    \frac{1}{t_\mathrm{grow}} = \frac{1}{E_c} \pdv{E_c}{t} \\
    = - \frac{1}{E_c}\left( \div{\vb{F}_c} + \left(\vb{u} + \vb{v}_s \right) \vdot \sigma_c \vdot \left( \vb{F}_c - \gamma_c \vb{u} E_c ) \right) \right).
\end{multline}
Note for this calculation, we have removed the term in Equation \ref{eqn:cr_energy} related to hadronic losses. Instead, this definition of $t_\mathrm{grow}$ allows us to identify how energy is transferred between the cosmic-rays and the gas. Figure \ref{fig:growth_time} uses the true change in average cosmic-ray energy density, including the hadronic losses.

In Figure \ref{fig:GrowthTimeVisual}, we also plot contours at $T=10^5\mathrm{K}$, highlighting any condensing clouds. Around these clouds, the growth time is mostly positive in all simulations, except \texttt{StrIND}. There are also plenty of red regions far away from the condensing clouds, as compressive waves wash over the diffuse gas. Those regions are where we expect turbulent acceleration to occur, with cosmic rays getting some net energy gain from large scale turbulence when they decouple from small scale fluctuations. But the fastest growth rates in Figure \ref{fig:GrowthTimeVisual} are generally near the edges of the condensing clouds. The cosmic rays initially get stuck in the clouds and experience the compression because the cooling time is short compared to the diffusion timescale. This process reaccelerates them just as a large scale turbulent eddy might. Then, on the diffusion timescale, the cosmic rays can escape the cold cloud before it is destroyed by heating or fragments into smaller clouds.

\begin{figure*}
    \centering
    \includegraphics[width=\linewidth]{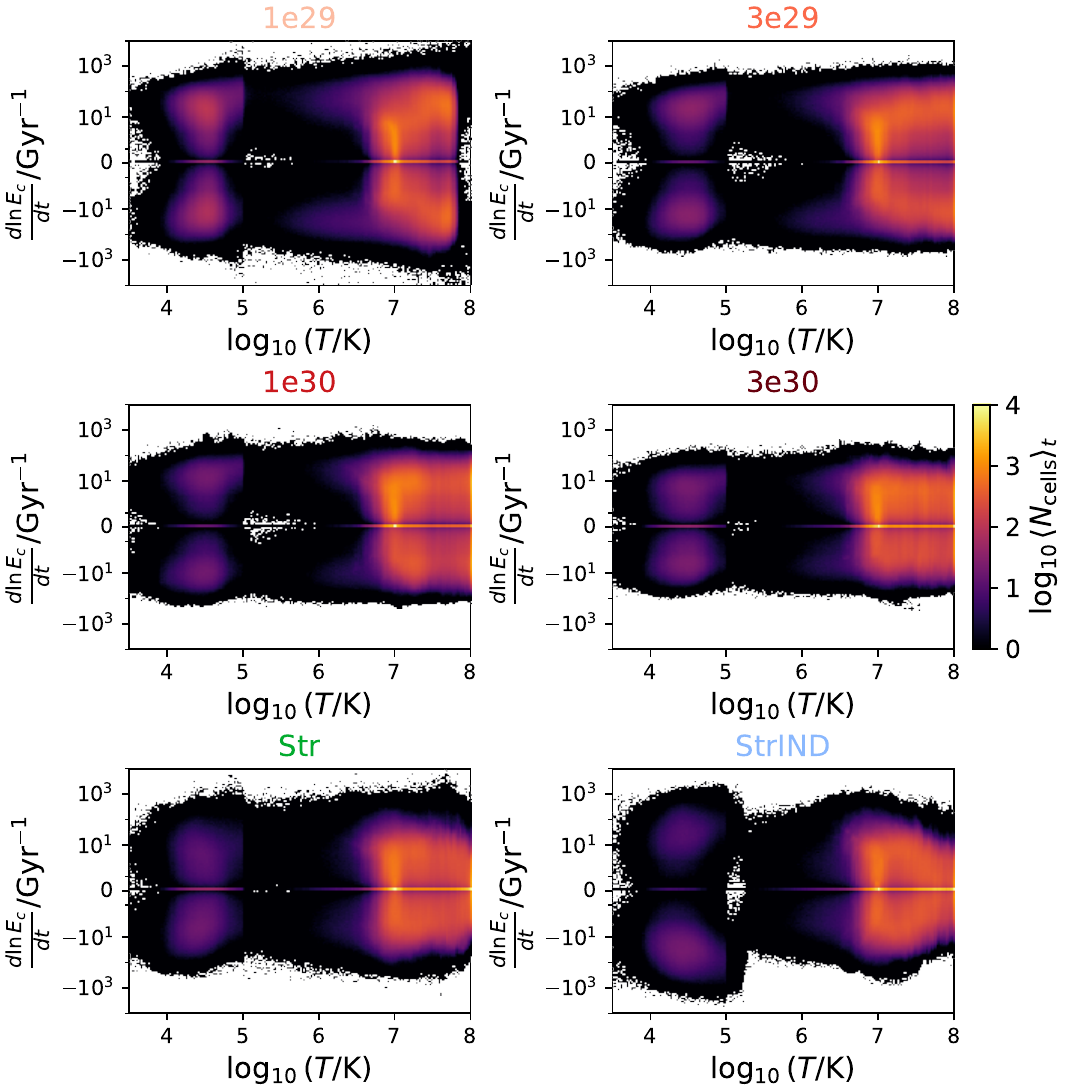}
    \caption{Average histograms of temperature and cosmic-ray energy-density-growth-time space. All cells over all time dumps (every $10\,\mathrm{Myr}$) are included here. Overall, there is symmetry between cosmic-ray energy density growth and loss. Most cells are at a temperature near $10^7\mathrm{K}$, with a smaller population of cells in colder gas at $3\times 10^{4}\mathrm{K}$. But taking the ratio of the upper and lower sections of this plot highlights the differences and net long term effects (see Figure \ref{fig:GrowthTimeHist_2}).}
    \label{fig:GrowthTimeHist_1}
\end{figure*}

\begin{figure*}
    \centering
    \includegraphics[width=\linewidth]{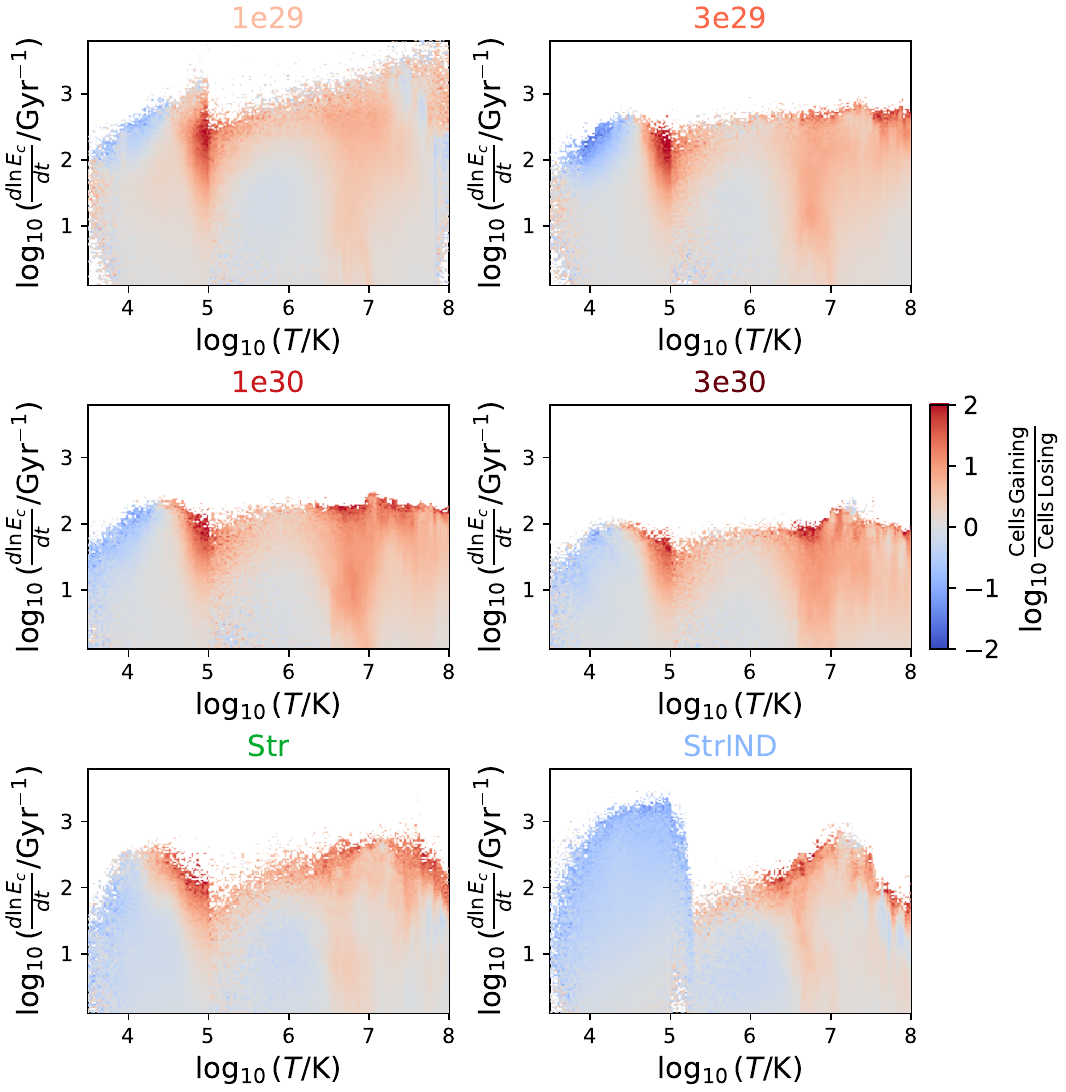}
    \caption{Ratio of the number of cells with cosmic-ray energy increase and the number of cells with cosmic-ray energy decrease, averaged over the simulation's runtime. Red regions show temperature regimes where cosmic rays are accelerated, and blue regions show where cosmic rays lose energy on average. The slight red region near $T \sim 10^7\,\mathrm{K}$ is due to traditional turbulent reacceleration, but the dark red regions at $10^5\mathrm{K}$ is due to condensing clouds of gas. This condensing cloud acceleration only disappears when the cosmic rays decouple from the gas via ion-neutral damping (\texttt{StrIND}) at the exact same temperature as the peak of the cooling curve.  }
    \label{fig:GrowthTimeHist_2}
\end{figure*}

\begin{figure*}
    \centering
    \includegraphics[width=\linewidth]{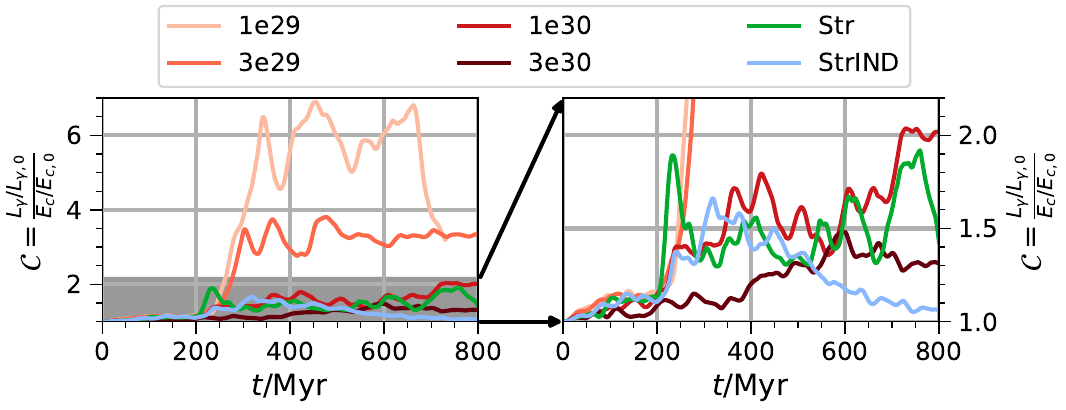}
    \caption{The left plot shows the evolution of the correlation coefficient $C$ we defined in Equation \ref{eqn:correlation_def} for each simulation. This coefficient divides out the effect of cosmic-ray acceleration, isolating the role of the correlation of cosmic-ray energy density and gas density in setting the $\gamma$-ray luminosity in the multiphase medium which forms after $t\sim 300\,\mathrm{Myr}$. The maximum effect the correlation has is a factor of $\sim 6$, in the slowest diffusion coefficient simulation. However, this does not account for the nearly two orders of magnitude increase illustrated in Figure \ref{fig:Lgamma}. instead, turbulent reacceleration is the source of that large increase in $\gamma$-ray luminosity. For the fast transport when turbulent reacceleration is inefficient (right plot), the correlation becomes the dominant factor in setting the $\gamma$-ray luminosity (compare to the right panel of Figure \ref{fig:Lgamma}). }
    \label{fig:correlation}
\end{figure*}

However, Figure \ref{fig:GrowthTimeVisual} is just one snapshot. In Figure \ref{fig:GrowthTimeHist_1}, we show how all cells over the entire simulation are distributed in temperature and $t_\mathrm{grow}$ space. The vertical symmetry of every simulation is readily apparent. In general, in the turbulent box, cosmic-ray energy is increasing just as often as it is decreasing. However, there are small deviations which explain the cosmic-ray energy growth apparent in the bottom plot of the left column of Figure \ref{fig:energies}. For example, all the diffusion simulations show a slightly bright vertical streak near $T \sim 10^7\mathrm{K}$. Most of the simulation volume is at this high temperature, so most cells occupy this region and most of them have minimal cosmic-ray energy growth or decay. The bright streak extending into the positive growth time regime illustrates the effect of turbulent reacceleration. This bright streak does not have a corresponding streak in the negative growth time regime, illustrating how cosmic rays are gaining energy on average.

The differences in the cold gas regime $(T\lesssim 10^5 \mathrm{K})$ are harder to see because fewer cells end up in this temperature regime. One apparent difference is in the \texttt{StrIND} simulation (bottom row, right column of Figure \ref{fig:GrowthTimeHist_1}). There are more cells in the negative growth rate regime for cold temperatures, showing that the increase in streaming velocity as a result of the change in ionization fraction leads to cosmic rays giving energy back to the thermal gas on average. The more interesting result is  in the diffusion simulations. The purple areas in the cold gas regime extend towards $T\sim 10^5 \mathrm{K}$ \textit{only in the positive growth time regime}. Therefore, there is reacceleration of cosmic rays occurring in this regime.

In Figure \ref{fig:GrowthTimeHist_2}, we show the ratio of the positive and negative growth time regimes for each simulation. These plots highlight regimes where cosmic rays are preferentially gaining energy (red) or preferentially losing energy (blue). It is important to note that regions higher in this plot are gaining energy at a faster rate --- the growth time gets as low as $\sim 1\, \mathrm{Myr}$ in some cases. 

All the simulations show some amount of turbulent reacceleration in the hot gas regime, with a red region slightly below $10^7\mathrm{K}$ but extending across at least two orders of magnitude in growth time. In general, there are about $10$ times as many cells with cosmic rays gaining energy in this temperature regime at any time dump. However, every simulation except \texttt{StrIND} shows a darker red region right at $10^5\, \mathrm{K}$, illustrating that on average there are $100$ times as many cells in this region gaining energy as opposed to losing it. 

This reacceleration at $10^5\mathrm{K}$ coincides with the peak of the cooling curve in our simulations. That temperature is the contour level used in Figure \ref{fig:GrowthTimeVisual}, so we know these cells are always around or near cooling clouds. However, given this is the peak of the cooling curve, we know gas at $T\sim 10^5 \,\mathrm{K}$ does not stay at that temperature. It is a transitory state as gas is cooling. However, that state is when compression and condensation is strongest, leading to a large and negative $\div {\vb{v}}$ which reaccelerates cosmic rays. \textbf{Not only does turbulent acceleration increase the cosmic-ray energy of our simulations, but there is a similar process driven by the formation of cold clouds in a multiphase medium}.

Now that we understand where the increase in cosmic-ray energy comes from, we can focus on the correlation of cosmic-rays and dense gas. To isolate the effect of dense gas-cosmic ray correlation, we divide out the evolution of the total cosmic ray energy density from the $\gamma$-ray luminosity. Those curves are shown in Figure \ref{fig:correlation} in the same style as Figure \ref{fig:Lgamma}. This quantity is equivalent to the correlation $\mathcal{C}$ defined in Equation \ref{eqn:correlation_def}. The correlation increases and peaks during the initial formation of cold gas and with the saturation of turbulence. However, the curves then level off at different values determined by the cosmic ray transport rate. The lowest diffusion coefficient run \texttt{1e29} has a correlation of $\sim 6$, whereas the fastest diffusion run \texttt{3e30} only reaches a maximum correlation of $\sim 1.5$. 

From Figure \ref{fig:correlation}, it is clear that while correlation plays a role in setting the luminosity, the two order of increase in $\gamma$-ray luminosity illustrated in Figure \ref{fig:Lgamma} must primarily result from the growth of cosmic-ray energy due to cooling-induced reacceleration (see Figure \ref{fig:growth_time}) and Figure \ref{fig:GrowthTimeHist_2}. The correlation only accounts for a small amount of this increase. 

The story is different when comparing the fast transport simulations (\texttt{3e30}, \texttt{Str}, and \texttt{StrIND}). On the right side of Figure \ref{fig:correlation}, the lines of correlation $\mathcal{C}$ are much closer to the amplitude in the change of the $\gamma$-ray luminosity. This result makes sense, since the total cosmic-ray energy barely changes for those simulations (see Figures \ref{fig:energies} and \ref{fig:growth_time}). Therefore, while reacceleration determines the growth of $\gamma$-ray luminosity when cosmic rays are transported slowly (left side of Figure \ref{fig:Lgamma}), it is the correlation between dense gas and cosmic rays which determines the $\gamma$-ray luminosity when cosmic rays are transported rapidly. 

\begin{figure*}
    \centering
    \includegraphics[width=0.48\linewidth]{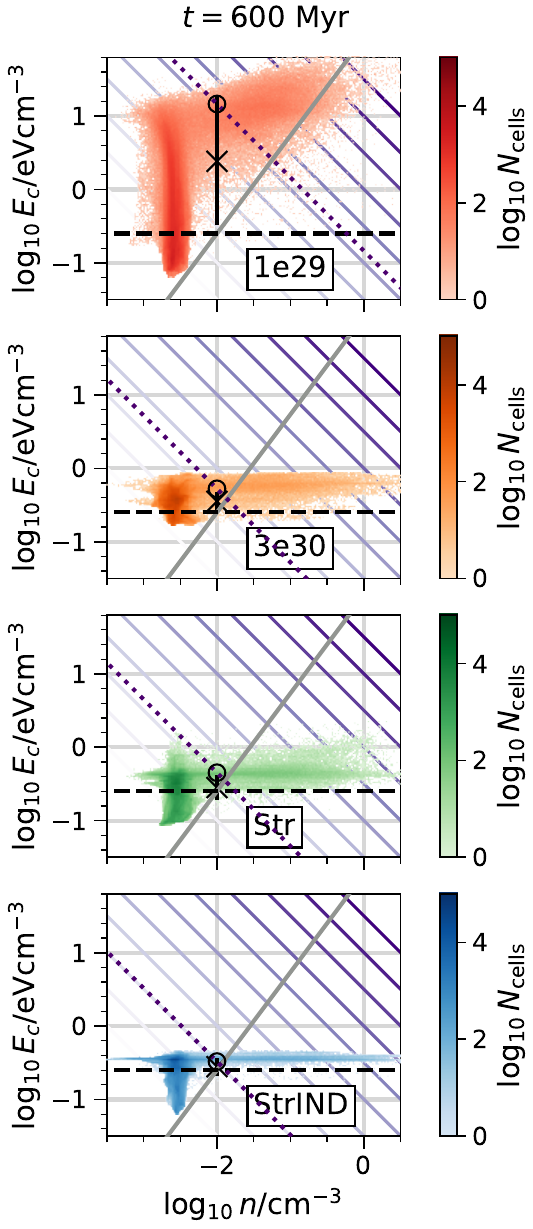}
    \includegraphics[width=0.48\linewidth]{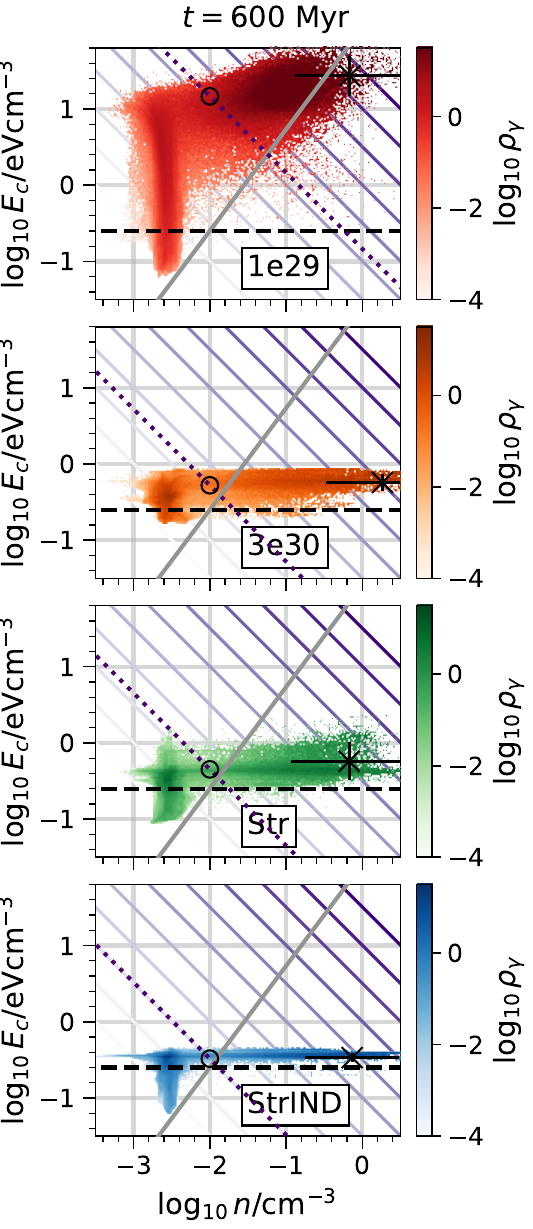}
    \caption{Cosmic-ray energy-gas density phase space distributions for the \texttt{1e29} (top row), \texttt{3e30} (second row), \texttt{Str} (third row), and \texttt{StrIND} (bottom row) simulations at $t=600\,\mathrm{Myr}$. For convenience, we plot the $\gamma$-ray emissivity contours and polytropic relationships, shown in Figure \ref{fig:Lgam_diagram}. We plot a dotted purple line along the average emissivity (proportional to the $\gamma$-ray luminosity) of each simulation at $t=600\,\mathrm{Myr}$, along with a black circle where assuming $\mathcal{C}=1$ puts the average cosmic-ray energy density, given the average gas density of $10^{-2}\mathrm{cm}^{-3}$ and average emissivity. The left column shows the raw distribution of computational cells, a volume weighted histogram. The right column re-weights the distribution using the $\gamma$-ray emissivity. Because the denser gas emits more $\gamma$-rays, that part of the distribution appears darker. For each weighting (volume and $\gamma$-ray emissivity), we calculate the average of $n$ and $E_c$, and mark this with an X. The error bars show the standard deviation of the average values. }
    \label{fig:Phasespace}
\end{figure*}

\subsection{Cosmic-Ray Energy Density in Dense Gas sets $\gamma$-ray Luminosity}\label{sec:results:densegas}

Now that we have an estimate for the effect of correlation on the $\gamma$-ray luminosity, we will take a closer look at the distribution of the simulations in $E_c$-$n$ phase space. As discussed in Section \ref{sec:phys:corr}, the distribution of gas in $E_c$-$n$ phase space determines the $\gamma$-ray luminosity. Therefore, all the changes due to cosmic-ray transport we have examined so far in Section \ref{sec:results} will be apparent in $E_c$-$n$ phase space. 

In Figure \ref{fig:Phasespace}, we illustrate how each simulation's volume is distributed in $E_c$-$n$ phase space at $t=600\,\mathrm{Myr}$.  For simulations \texttt{1e29}, \texttt{3e30}, \texttt{Str}, and \texttt{StrIND}, we show how many cells are in each region of cosmic-ray energy density vs. gas density phase space in the left column of Figure \ref{fig:Phasespace}. We also plot the contour lines of $\gamma$-ray emissivity and the advective vs. diffusive transport polytropic indices from Figure \ref{fig:Lgam_diagram}. The distributions change significantly across the different diffusion rates and transport mechanisms. 

All the simulations are predominantly diffuse gas $(n\lesssim10^{-2}\mathrm{cm}^{-3})$ and not dense gas, as we saw in Figure \ref{fig:thermodynamics}. However, in the case where reacceleration is important (the \texttt{1e29} simulation in the top plot), the cosmic-ray energy density is much higher in the dense gas. Similarly, the cosmic-ray energy density stays even lower in the dense gas of the \texttt{StrIND} simulation than in the \texttt{3e30} and \texttt{Str} simulations, as a result of the increase in the Alfv\'en speed due to the change in ionization. 

Each plot has an X marking the mean values of cosmic-ray energy density and gas density. The error bar shows one standard deviation in $\langle E_c \rangle$, but we do not show the error bar in $\langle n \rangle$. Looking back at the marginal density distribution in Figure \ref{fig:thermodynamics} illustrates the density is bimodal, so using a single standard deviation is inaccurate. Calculating it leads to an error bar stretching the entire span of density values, from $10^{-4} \mathrm{cm}^{-3}$ to $1\,\mathrm{cm}^{-3}$. However, since mass is conserved in the simulations, this average density of $10^{-2}$ stays constant during the entire simulation. The final distribution of densities therefore appears ``unconstrained'' because of the large range in density values, but the average value is extremely constrained. 

Each plot also has a dotted purple line. This line traces along the average emissivity (proportional to the total luminosity) for each simulation at the given time. The separation of this line from the location of the average density and average cosmic-ray energy is captured by the correlation factor $\mathcal{C}$ (see Figure \ref{fig:correlation}). 

We also plot a black circle (or O), marking where a $\gamma$-ray observation of our simulation, combined with the average density $10^{-2}\mathrm{cm}^{-3}$, would place the average cosmic-ray energy density. The disagreement of the X and O points in the left column is the correlation $\mathcal{C}$. 

The left column of Figure \ref{fig:Phasespace} shows that a single measurement of $\gamma$-ray luminosity is not a reliable measure of the distribution of gas in $E_c$-$n$ phase space. This luminosity will fall in between the dense gas and the diffuse gas, and attempting to calculate $\langle E_c\rangle$ from that luminosity and $\langle n \rangle$ without considering this distribution could cause a severe overestimate of the cosmic-ray energy density. The overestimate depends on the cosmic-ray transport in both the hot and cold phases of the plasma, and is determined by the correlation factor $\mathcal{C}$. Considering the plots of the faster transport simulations, where $\mathcal{C}$ is close to $1$, it is apparent that a minimal amount of gas actually has the cosmic-ray energy density and gas density estimated by the volume averages. Instead, the filling factor of the dense gas causes the diffuse component to be at a lower density, such that when averaged with the higher density gas the expected value lands in between the two phases.

In the right column of Figure \ref{fig:Phasespace}, we show the same histograms, reweighted by the $\gamma$-ray emissivity of each cell. We then calculate the expectation value or mean value of the cosmic-ray energy density and gas density with these weights to find which part of the distribution contributes the most to the final $\gamma$-ray luminosity. Those points are marked with an X and have error bars corresponding to one standard deviation of each variable (cosmic-ray energy density and gas density). Across all transport mechanisms, the dense gas dominates the $\gamma$-ray luminosity, despite it not being volume filling  (illustrated by there being lower values of $N_\mathrm{cells}$ in the dense gas regime in the histograms of the left column of Figure \ref{fig:Phasespace}). 

We again plot a dark purple, dotted line corresponding to the average emissivity. The weighted average of cosmic-ray energy density and gas density is much higher than the average emissivity. Effectively, the picture is one where the cosmic-ray energy density in this dense gas pulls the average luminosity upward.

We also plot the black circle marking a mock observational result. Across all transport mechanisms, the observed value of $E_c$ is more in line with the emissivity weighted average cosmic-ray energy density than the volume weighted average cosmic-ray energy density. \textbf{Therefore, if the diffuse gas of the CGM or ICM is multiphase and has some dense gas, the $\gamma$-ray luminosity probes the average cosmic-ray energy density in the dense gas --- not the cosmic-ray energy density in the diffuse gas.}

It is important to note that the $\gamma$-ray-emissivity weighted mean values of gas density and cosmic-ray energy density are different from the mean values used in calculating the $\gamma$-ray luminosity. Returning to Equation \ref{eqn:luminosity}, we can see the weighting is purely based on volume. Therefore, taking mean values of gas density and cosmic ray energy on the left side column of Figure \ref{fig:Phasespace} is still a reasonable way to estimate the $\gamma$-ray luminosity. But the right side of Figure \ref{fig:Phasespace} highlights how it is important to consider any (possibly unresolved) cold and dense gas when observing the $\gamma$-ray emission from diffuse gas. If the medium is multiphase with small volumes of dense gas, then those cold, dense clumps could dominate the $\gamma$-ray emission.

\begin{figure}
    \centering
    \includegraphics[width=\linewidth]{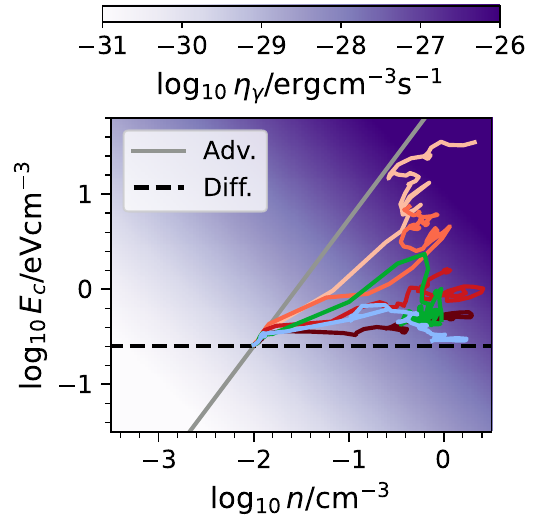}
    \caption{Following up on the calculation of the $\eta_\gamma$-weighted centroids in Figure \ref{fig:Phasespace}, we plot the movement of the emissivity weighted centroids (X markers in the right side of Figure \ref{fig:Phasespace}) over time for each simulation. The diffusion and advection polytropic laws intersect at the initial state of each simulation. Each curve follows a spectrum of slopes, with slower cosmic-ray transport leading to movement close to the advection polytropic law. Overall, transport dictates what the cosmic ray energy density is in dense gas, and this significantly adjusts the total $\gamma$-ray output. }
    \label{fig:LgamEvolution}
\end{figure}

We also can examine how the $\gamma$-ray-emissivity weighted phase space centroids in the right column of Figure \ref{fig:Phasespace} evolve through time. Figure \ref{fig:LgamEvolution} shows how the weighted average density and cosmic-ray energy density evolve with time in $E_c$-$n$ phase space. Overall, the centroids follow the simple polytropic laws but have key deviations. Initially, they all follow the advection curve. Once diffusion takes over, the lines move to the right, with lower diffusion coefficients staying closer to the advection case. The polytropic law is also apparent in the simple streaming case without ion-neutral damping (simulation \texttt{Str}, green line in Figure \ref{fig:LgamEvolution}), which follows a $2/3$ polytropic index before cold gas forms. That index is expected for streaming at the Alfv\'en speed and results from the $\sqrt{\rho}$ in the Alfv\'en speed when examining a steady state system. 

Note, we do not plot the standard deviations in Figure \ref{fig:LgamEvolution}. The dense gas ends up with a variety of densities in the \texttt{3e30}, \texttt{Str}, and \texttt{StrIND} simulations, but a significantly lower emissivity weighted cosmic-energy density since the cosmic rays can easily escape the dense regions in these simulations.

\section{Discussion}\label{sec:disc}

In this section, we discuss the implications of our results. In particular, we discuss how our examination of the correlation factor $\mathcal{C}$ can be used in estimating the average cosmic-ray energy density in diffuse gas and how turbulent reacceleration by condensing clouds could have a significant impact on the dynamics of turbulence in diffuse media like the CGM or ICM.

As mentioned in Section \ref{sec:intro}, observations of $\gamma$-rays from environments like the ICM and CGM provide one of the few probes of the cosmic-ray energy density $E_c$ in those environments. While we at least have some chance at direct probes of the low-energy cosmic rays $\sim 1-10 \,\mathrm{GeV}$ in the solar neighborhood, we lack a direct probe in these diffuse, large scale environments. Therefore, it is important that we understand all the assumptions we make about the diffuse medium and how those assumptions introduce error into derived quantities like an average cosmic-ray energy density.  

Previous works have focused on the full structure of galaxy clusters, allowing for a direct comparison with $\gamma$-ray observations \citep{2010Pinzke,2022A&A...665A.129B}. The end conclusion from most of these simulations is to continue decreasing the average cosmic-ray energy density, so that the estimate $\gamma$-ray emission is within a reasonable value. That is still necessary, but our work addresses another possible solution: that fast cosmic-ray transport is necessary to keep $\gamma$-ray emission low. In particular, our simulations illustrate that cosmic rays need to decouple from dense gas regions in order to have a low $\gamma$-ray luminosity.

Obviously, the simplest estimate of cosmic-ray energy density is the ratio of the observed $\gamma$-ray luminosity to the average gas density (ideally measured with another observable quantity besides $\gamma$-ray luminosity). That estimation method is shown in Equation \ref{eqn:simple}. But as we have illustrated, the turbulent multiphase nature of astrophysical plasmas limits the accuracy of this method. A good first correction is to consider the correlation of cosmic rays with dense gas, and include this as a factor $\mathcal{C}$. This correction can be estimated from an assumed distribution of gas in $E_c$-$n$ phase space. We have already shown the value of $\mathcal{C}$ can range from $6$-$1.5$, depending on the cosmic-ray transport in the medium. For more complex simulations with more realistic cosmic-ray transport, the value could be different. 

We also illustrated how the $\gamma$-ray luminosity could be increased as a result of turbulent reacceleration \citep{1988SvAL...14..255P}. This mechanism caused the largest change in $\gamma$-ray luminosity (see Figures \ref{fig:Lgamma} and \ref{fig:growth_time}), and was an even stronger effect than the correlation. Together, the correlation and the turbulent reacceleration suggest that the $\gamma$-ray luminosity observations are probing the cosmic-ray energy density in denser gas components. This conclusion derives from the definition of the $\gamma$-emissivity; a denser gas with higher cosmic-ray energy density will emit more $\gamma$-rays (see the right column of Figure \ref{fig:Phasespace}). 

There is a nuance to turbulent reacceleration which we have not discussed. The reacceleration only matters to $\gamma$-ray luminosity if the cosmic rays gaining energy exceed the energy threshold for hadronic interaction $(E_\mathrm{thr} = 1.282 \mathrm{GeV})$. Additionally, if reacceleration pushes cosmic rays to higher energies such that the emission is outside observable bands of $\gamma$-ray emission, then the reacceleration will not impact the $\gamma$-ray luminosity. Overall, the evolution of the cosmic-ray distribution function under the influence of turbulent reacceleration would have a significant impact on the evolution of the $\gamma$-ray luminosity. However, we are unable to address that process. Our simulations do not evolve separate bands of the cosmic-ray distribution function, instead they evolve the total cosmic-ray energy density. 

The details of the turbulent reacceleration we observe require further study. In particular, with regard to the evolution of the cosmic-ray distribution and the reacceleration by condensing cold clouds. But our results show that this process exacerbates the $\gamma$-ray luminosity problem identified in other works. If cosmic-ray transport does not become fast in colder, denser gas, then the $\gamma$-ray luminosity becomes too large to agree with observations. Turbulent acceleration further increases the cosmic-ray energy density in dense gas, increasing the $\gamma$-ray luminosity more. 

The final point of discussion regards how the cosmic-ray energy growth time in our simulations (Figure \ref{fig:growth_time})  disagreed with the analytic estimates in \cite{1988SvAL...14..255P,2022ApJ...941...65B,2023ApJ...955...64B}. We found the cause of disagreement by looking at where in the simulation cosmic-ray energy was increasing the most. We found that $T\sim 10^5 \mathrm{K}$ gas was the dominant contributor to cosmic-ray reacceleration in Figure \ref{fig:GrowthTimeHist_2}. That temperature corresponds to cooling, condensing gas clouds (see Figure \ref{fig:GrowthTimeVisual}) as that temperature is where our simple cooling curve peaks (see Equation \ref{eqn:Cooling}). However, if the cosmic rays effectively decouple from the gas at the same temperature (the case of simulation \texttt{StrIND}), then the reacceleration via condensing cold clouds does not occur. 

In reality, the cooling function and cosmic-ray transport in multiphase plasma is more complex than what we implement here. It is likely that the reacceleration via condensing clouds could still occur, depending on how changes in ionization correspond to changes in the cooling rate. The peak in the cooling curve for solar metallicity gas is near $T\sim 10^5 \mathrm{K}$, but ionization is more likely to shift at $T\sim 10^4\mathrm{K}$. Therefore, one could end up with models where cosmic rays are coupled to regions of condensing warm ionized medium (WIM) surrounding regions of warm neutral medium (WNM). The cosmic-ray reacceleration would occur in this shell of WIM, but the cosmic rays could decouple from the WNM and cold neutral medium (CNM) of dense clouds of gas. In the future, we will examine the collapse of an individual cloud from diffuse gas in ICM, CGM, and ISM conditions, including more realistic and complex physical descriptions of cooling and cosmic-ray transport. Without those detailed simulations, it is difficult to conclude that cosmic-ray reacceleration by condensing clouds actually occurs in any astrophysical medium. However, that mechanism does describe the reacceleration in our simulations, which deviated from the analytic estimates and simulations of single-phase turbulent media.

\section{Conclusion}\label{sec:conc}

In this work, we presented an in-depth look at how cosmic-ray transport influences the $\gamma$-ray luminosity of diffuse gas. Overall, we have illustrated how interpretation of $\gamma$-ray luminosity observations of diffuse gas requires several assumptions about cosmic-ray transport, the interaction of cosmic rays with turbulence, and the correlation of cosmic rays with dense gas.  Below, we provide a list of the individual key results from this paper, in no particular order of importance. We also point to the figures which illustrate each conclusion.
\begin{itemize}
    \item Cosmic-ray transport needs to be rapid in diffuse gas on large length scales, with either a large diffusion coefficient $\gtrsim 10^{30}\mathrm{cm}^2 \mathrm{s}^{-1}$, or via streaming at the Alfv\'en speed. Otherwise, the $\gamma$-ray luminosity will be extremely large. Specifically, fast transport reduces $\gamma$-ray luminosity by a factor of at least 100 compared to slower transport. (See Figure \ref{fig:Lgamma}) 
    \item Turbulent reacceleration in diffuse gas will increase the average cosmic-ray energy, thereby driving the $\gamma$-ray luminosity up. This effect shows it is important to check the efficiency of turbulent reacceleration as well as reacceleration by condensing clouds. (See Figure \ref{fig:growth_time})
    \item We show that cosmic-ray energization by condensing, cooling clouds could be more effective than single-phase turbulent reacceleration suggested by \cite{1988SvAL...14..255P}. Additionally, it operates at length scales associated with the cooling processes, instead of only being determined by cosmic-ray transport. The cosmic-ray transport can stifle this cold cloud reacceleration if the cosmic rays decouple from the gas at the same temperature as the phase transition. (See Figures \ref{fig:GrowthTimeVisual},\ref{fig:GrowthTimeHist_1}, and \ref{fig:GrowthTimeHist_2})
    \item The correlation of cosmic rays with dense gas causes a factor of $1.5$-$6$ difference in estimates of the average cosmic-ray energy density. This correlation itself is a subdominant factor in determining the boost in $\gamma$-ray luminosity. Instead, it is the turbulent and cooling-induced reacceleration that drives the level of $\gamma$-ray production. (See Figures \ref{fig:render}, \ref{fig:Phasespace} and \ref{fig:correlation})
    \item Even when cold, denser gas is not volume filling, it is the dominant factor in setting the $\gamma$-ray luminosity. This conclusion results directly from the dependence of $\gamma$-ray luminosity on the local gas density and cosmic-ray energy density. However, it implies that estimates of average cosmic-ray energy density from observations of $\gamma$-ray luminosity are estimates of the cosmic-ray energy density in any dense gas clumps in the medium, as opposed to the cosmic-ray energy density in the diffuse component. (See Figures \ref{fig:Phasespace} and \ref{fig:LgamEvolution})
\end{itemize}

\begin{acknowledgments}
The authors would like to thank Karol Fulat, Ka Wai Ho, Mohan Richter-Addo, Aaron Tran, Bindesh Tripathi, and Ka Ho Yuen for their insight and helpful conversations which improved this work. 

We appreciate computational time on NCSU Delta provided by NSF ACCESS allocation PHY240310. 

RH and EZ greatly appreciate funding from  NASA FINESST grant No. 80NSSC22K1749 and NSF grant AST-2007323 which supported this work.
MR acknowledges support from the National Science Foundation Collaborative Research Grant NSF AST-2009227.
This work was performed in part at the Aspen Center for Physics (ACP) during the ``{\it Cosmic Ray Feedback in Galaxies and Galaxy Clusters}" summer program. The ACP is supported by the National Science Foundation grant PHY-2210452, and by grants from the Simons Foundation (1161654, Troyer) and Alfred P. Sloan Foundation (G-2024-22395). 
This research was supported in part by grant NSF PHY-2309135 to the Kavli Institute for Theoretical Physics (KITP).
\end{acknowledgments}

\vspace{5mm}

\software{Athena++ \citep{2020Stone,2018Jiang}, MatPlotLib \citep{2007Matplotlib}, NumPy \citep{2011NumPy,2020NumPy}, AstroPy \citep{2013AstroPy,2018Astropy}}


\bibliography{sample701}{}
\bibliographystyle{aasjournalv7}





\end{document}